\documentclass[%
 reprint,
amsmath,amssymb,
aps,
longbibliography,
]{revtex4-2}

\usepackage[pdftex]{graphicx}

\graphicspath{
{./Figures/},
}

\usepackage{chemformula} 
\usepackage{diagbox}
\usepackage{braket}
\usepackage{dcolumn}
\usepackage{bm}
\usepackage{booktabs}
\usepackage{threeparttable}

\begin{document}

\preprint{APS/123-QED}

\title{Mathematical crystal chemistry I\hspace{-1.2pt}I: Random search for ionic crystals and analysis on oxide crystals registered in ICSD}

\author{Ryotaro Koshoji}
\altaffiliation[Present address: ]{Institute for Solid State Physics, The University of Tokyo, Kashiwa 277-8581, Japan}
\email{cosaji@issp.u-tokyo.ac.jp}
\affiliation{National Institute of Advanced Industrial Science and Technology, Nagoya, 463-8560, Japan}

\date{\today}

\begin{abstract}

\noindent
Mathematical crystal chemistry views crystal structures as the optimal solutions of mathematical optimization problem formalizing inorganic structural chemistry.
This paper introduces the minimum and maximum atomic radii depending on the types of geometrical constraints, extending the concept of effective atomic sizes.
These radii define permissible interatomic distances instead of interatomic forces, constraining feasible types and connections of coordination polyhedra.
The definition shows the aspect that crystal structures are packings of atomic spheres.
Additionally, creatability functions for geometrical constraints, which give a choice of creatable types of geometrical constraints depending on the spatial order of atoms, are implemented to guide randomly generated structures toward optimal solutions.
The framework identifies unique optimal solutions corresponding to the structures of spinel, pyrochlore ($\alpha$ and $\beta$), pyroxene, quadruple perovskite, cuprate superconductor \ch{YBa2Cu3O_{7-$x$}}, and iron-based superconductor \ch{LaFeAsO}.
Notably, up to $95\%$ of oxide crystal structure types in Inorganic Crystal Structure Database align with the optimal solutions preserving experimental structures despite the discretized feasible atomic radii.
These findings highlight the role of mathematical optimization problem as a theoretical foundation for mathematical crystal chemistry, enabling efficient structure prediction.

\end{abstract}

\maketitle

\section{introduction}

\begin{figure*}
\centering
\includegraphics[width=1.9\columnwidth]{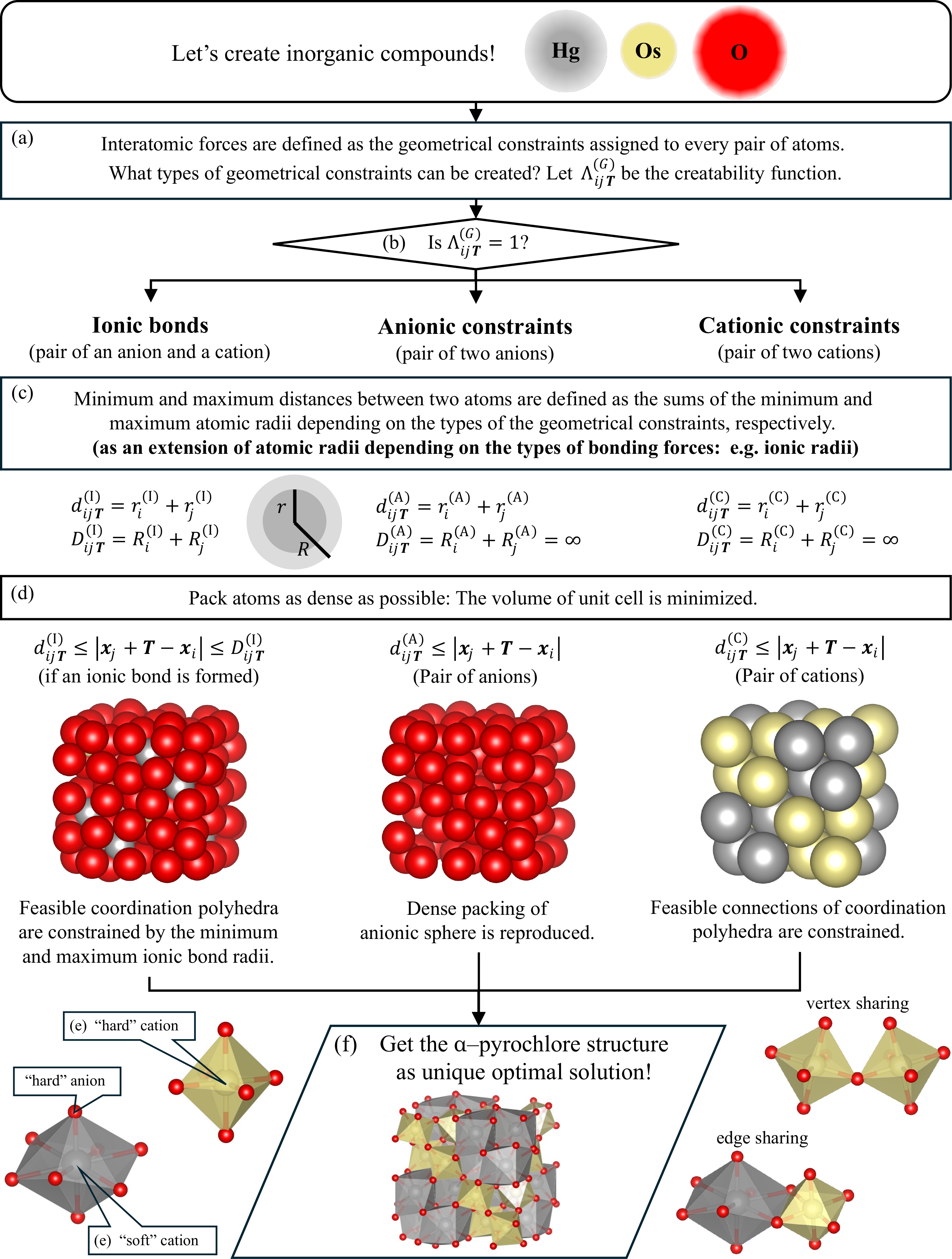}
\caption{
The concept of mathematical crystal chemistry.
(a) The creatability function $\Lambda_{ij \bm{T}}^{\left( G \right)}$ is defined in Eq.~\eqref{eq:constraint_creatability}.
(b) The creatability functions are evaluated in the mathematical optimization problem give by Eq.~\eqref{eq:equation_of_crystal_structures}.
(c) The feasible interatomic distance is defined in Eq.~\eqref{eq:geometrical_constraints}, and the minimum and maximum interatomic distances are defined in Eqs.~\eqref{eq:minimum_distance} and \eqref{eq:maximum_distance}, respectively.
(d) The objective function of the mathematical optimization problem is the volume of unit cell as given in Eq.~\eqref{eq:equation_of_crystal_structures}.
(e) Flexible atomic sphere is discussed in Appendix~\ref{sec:discretized_feasible_atomic_radii}.
(f) A wide variety of crystal structures identified as unique optimal solutions are discussed in Sec.~\ref{sec:result_and_discussion}.
}
\label{fig:concept}
\end{figure*}

Inorganic Crystal Structure Database (ICSD)~\cite{doi:10.1021/ci00038a003} contains around 250000 experimental inorganic structures as of 2025, but they only covers a tiny portion of the possible inorganic compounds considering the infinite combinations of the chemical elements across the periodic table.
This is due to the traditional trial-and-error approach to discover high-performance functional materials, being time-consuming and labor-intensive.
The current state-of-the-art approaches to accelerate the discovery of unknown materials is the computational design of materials, but it remains challenging due to the sheer size of freedom such as chemical compositions, unit cells, and the atomic arrangements.

The standard approaches for crystal structure prediction mainly combine density functional theory calculations with structure searching algorithms such as random search~\cite{PhysRevLett.97.045504, Pickard_2011}, genetic algorithms~\cite{10.1039/9781788010122, doi:10.1063/1.2210932, Liu2021}, and particle-swarm optimization~\cite{PhysRevB.82.094116, WANG2016406}.
These methods lead to a number of major discoveries such as the high pressure stabilized superconductor \ch{LaH10}~\cite{PhysRevLett.119.107001, Liu6990, doi:10.1002/anie.201709970, Drozdov2019, PhysRevLett.122.027001, doi:10.1063/1.5128736, https://doi.org/10.1002/adma.202006832}.
However, they suffer from their limited applicability to complex structures with large number of atoms of several kinds in the unit cells, because they are based on the direct sampling over the entire energy surface; the large computational cost of \textit{ab-initio} simulations causes a problem of insufficient sampling for a large system.

In recent years, machine learning methods such as graph neural networks representing a structure of compounds with nodes and edges~\cite{PhysRevLett.120.145301, PhysRevMaterials.4.063801, doi:10.1021/acs.chemmater.9b01294, Jiang2021, Cheng2021, Fung2021, Choudhary2021, Reiser2022} have been introduced to provide a faster way to predict properties without using density functional methods.
The graph neural network appoaches can be applied to crystal structure prediction with combining symmetry constraints~\cite{Li2023}, graph theory techniques to generate initial structures~\cite{Xiao2023}, or the structure search algorithm such as particle-swarm optimization~\cite{Cheng2022, Merchant2023}.
Diffusion-based generative model is also presented for tackling inverse design tasks under desired property constraints~\cite{Zeni2025}.
On the other hand, biased symmetry-adapted artificial intelligence algorithms successfully predicted complex crystal structures such as the garnet structure~\cite{10.1063/5.0074677}, because sampling with space group significantly simplifies the problem of crystal structure prediction by generating more promising initial structures than random sampling.
The integer programming approach also found the garnet structure as the global optimum with the space group constraint to narrow down the combinations of the discretized atom sites in the fixed cubic unit cell, where the total energy is calculated by classical two-body potentials so that the structural optimization problem is converted into the integer programming formalism~\cite{Gusev2023}.

The previous study on mathematical crystal chemistry~\cite{PhysRevMaterials.8.113801} proposed a mathematical optimization problem formalizing the empirical rules of inorganic structural chemistry~\cite{InorganicStructuralChemistry, StructuralInorganicChemistry, doi:10.1021/ja01379a006}.
The structural aspects such as the sizes of atoms, the coordination polyhedra, and connection of polyhedra, are formalized as the objective or constraint functions.
For example, a chemical bond is defined to be the inequality constraints on the distances between atoms, and the objective function is given by the volume of the unit cell to reproduce the packing of atomic spheres.
The mathematical model successfully made a wide variety of crystal structures such as spinel and $\alpha$-pyrochlore structures.
Besides, the number of optimal solutions was found to be much smaller than the number of local minima in the total energy by \textit{ab-initio} simulations.

This study provides a more direct mathematical formalization of inorganic structural chemistry based on the previous study~\cite{PhysRevMaterials.8.113801} in order that the concepts of mathematical crystal chemistry should be clarified as summarized in Fig.~\ref{fig:concept}.
The previous study~\cite{PhysRevMaterials.8.113801} explicitly constrains feasible connections of coordination polyhedra by maximum number of common bridging anions, but they should be constrained by minimum distances between cations, because inorganic structural chemistry considers that the connection of polyhedra results from the electrostatic stability as highly charged cations prefer the vertex sharing to edge or face sharing of coordination polyhedra to lengthen the distances between cations.
Note that mathematical crystal chemistry defines interatomic forces as geometrical constraints on the interatomic distances.
Besides, this study explicitly defines the minimum and maximum atomic radii depending on the type of geometrical constraints as the extention of the effective sizes of atoms depending on the types of bonding forces.
Furthermore, the rule to assign one out of the types of geometrical constraints to every pair of atoms is mathematically formulated by introducing the creatability function $\Lambda_{ij \bm{T}} ^{\left( G \right)}$ similar to activation functions in neural networks, resulting in unifying the paired optimization problem derived in the previous study~\cite{PhysRevMaterials.8.113801}.
The unified formalization is also shown to have the capability of discovering a wide variety of ionic crystal structures as optimal solutions from randomly generated structures.
Finally, oxide crystals registered in ICSD are analyzed whether they can be the optimal solutions preserving the experimental structures by assigning suitable coordination features to atoms.
As a result, $78\%$, $12\%$, and $4\%$ of oxide crystals have isotypic, homeotypic, or isopointal optimal structures, respectively, despite the discretized feasible atomic radii.
The successful results shows that mathematical optimization problem can be the theoretical foundation of mathematical crystal chemistry to explore the crystal structure prototypes with small computation.

This paper is organized as follows: In Section \ref{sec:mathematical_crystal_chemistry}, the theoretical foundation of mathematical crystal chemistry is discussed.
In section \ref{sec:methods}, the computational aspect is described.
In section \ref{sec:result_and_discussion}, the application results are shown.
In Section \ref{sec:conclusion}, the comprehensive summary is provided.

\section{Mathematical crystal chemistry}
\label{sec:mathematical_crystal_chemistry}

\begin{figure}
\centering
\includegraphics[width=1.0\columnwidth]{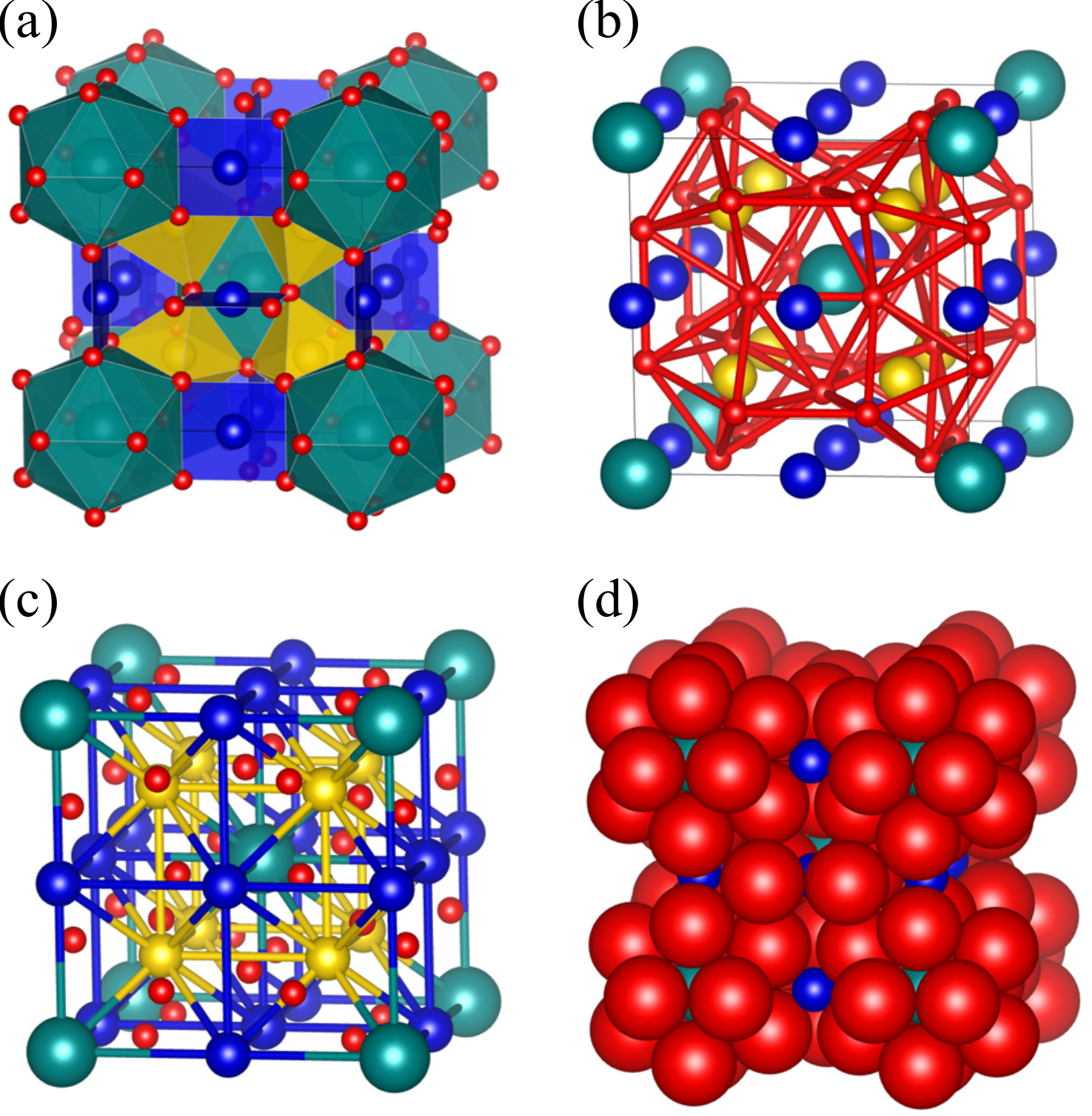}
\caption{
The different structural aspects of ionic crystals depending on the type of the geometrical constraints that correspond to the types of interatomic forces.
In this study, all the figures of crystal structures are drawn by VESTA~\cite{Momma:db5098}.
(a) An ionic bond can be formed between a pair of an anion and a cation unless their interatomic distance will be infeasible when it is formed.
(b) Anionic constraints are formed between every pair of anions.
(c) A cationic constraint is formed between a pair of metal atoms if they have common bridging anions.
(d) Packing of ionic spheres. If the sizes of cations are small enough, they are placed in the tetrahedral or octahedral site of the densest packings of oxide ions.
}
\label{fig:geometrical_constraints}
\end{figure}
\begin{figure*}
\centering
\includegraphics[width=2.0\columnwidth]{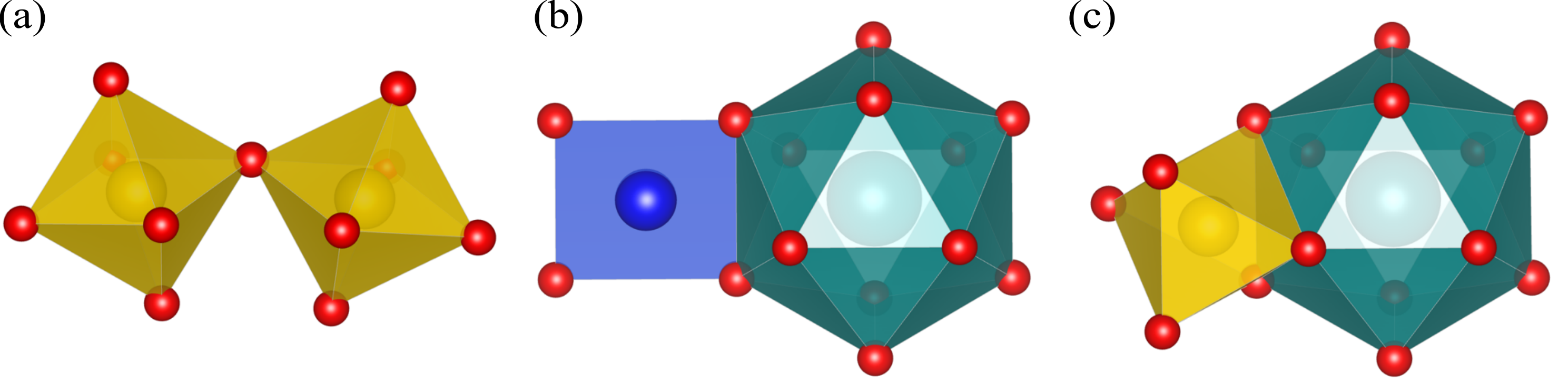}
\caption{
Connection of polyhedra in the quadruple perovskite structure of \ch{CaCu3Fe4O12}.
Calcium, copper, iron, and oxygen atoms are shown as darkcyan, blue, gold, and red balls, respectively.
All the minimum distances between cations are $2 C^{\left( \mathrm{Qd} \right)}$, but the feasible connection of two polyhedra depends on the sizes of the two polyhedra.
(a) Vertex sharing of two octahedra. The edge or face sharings are infeasible due to the large minimum distances between central cations.
(b) Edge sharing of square and regular icosahedron. The edge sharing is feasible due to the large size of regular icosahedron.
(c) Face sharing of octahedron and regular icosahedron. The face sharing is feasible due to the large size of regular icosahedron.
}
\label{fig:linked_polyhedra}
\end{figure*}

Interatomic forces stabilize a crystal structure by keeping all the interatomic distances within the stable ranges.
Accordingly, mathematical crystal chemistry defines an interatomic force as the geometrical constraint $G$ that constrains the feasible interatomic distance as
\begin{equation}
d_{ij \bm{T}}^{\left(G \right)} \le x_{\sigma} \le D_{ij \bm{T}}^{\left(G \right)},
\label{eq:geometrical_constraints}
\end{equation}
where $d_{ij \bm{T}}^{\left(G \right)}$ and $D_{ij \bm{T}}^{\left(G \right)}$ are the minimum and maximum distances between atom $i$ and atom $j$ in lattice $\bm{T}$, respectively, and $x_{\sigma}$ is given by
\begin{equation}
x_{\sigma} = \left| \bm{x}_j + \bm{T} - \bm{x}_i \right|,
\end{equation}
with $\bm{x}_i$ being the cartesian coordinate of atom $i$ in the original cell, $\bm{T}$ being the lattice vector, and $\sigma$ being the abbreviated index of $ij \bm{T}$.
In the following discussion, the atom $j$ in lattice $\bm{T}$ is called atom $j \bm{T}$.
Since randomly generated structures are too difficult to satisfy all the constraints formed between every pair of atoms, a geometrical constraint is relaxed to be
\begin{equation}
\left( 1 - \varepsilon \right) d_{\sigma}^{\left(G \right)} \le x_{\sigma} \le \left( 1 + \varepsilon \right) D_{\sigma}^{\left(G \right)},
\label{eq:relaxed_geometrical_constraints}
\end{equation}
where $\varepsilon$ is an error rate.
The geometrical constraints are classified according to the types of interatomic forces based on inorganic structural chemistry, and one out of the types of geometrical constraints is assigned to each pair of atoms.

Figure~\ref{fig:geometrical_constraints}(a), \ref{fig:geometrical_constraints}(b), and \ref{fig:geometrical_constraints}(c) show the ionic bonds $\mathrm{I}$, anionic constraints $\mathrm{A}$, and cationic constraints $\mathrm{C}$ formed in the quadruple perovskite structure, respectively.
The three types of the geometrical constraints are formed between pairs of atoms depending on the formal charges, and they show the different structural aspects of ionic crystals.
In addition to the three kinds of constraints, non-ionic bond $\mathrm{N_I}$ and metallic constraint $\mathrm{M}$ are also defined as the types of the geometrical constraints in this study.
Figure~\ref{fig:geometrical_constraints}(d) shows another aspect that ionic crystals are packings of ionic spheres, in fact, oxide ions in crystals with small cations tend to constitute the densest packings of spheres~\cite{10.2307/20159940}.

Since inorganic structural chemistry estimates a chemical bonding distance as the sum of atomic radii, mathematical crystal chemistry defines the minimum and maximum interatomic distances as
\begin{align}
d_{ij \bm{T}}^{\left(G \right)} &= r_i ^{\left(G \right)} + r_j ^{\left(G \right)}, \label{eq:minimum_distance} \\
D_{ij \bm{T}}^{\left(G \right)} &= R_i ^{\left(G \right)} + R_j ^{\left(G \right)}, \label{eq:maximum_distance}
\end{align}
where $r ^{\left(G \right)}$ and $R ^{\left(G \right)}$ are the minimum and maximum atomic radii depending on the type of the geometrical constraint $G$, respectively.
The several kinds of the atomic radii, which determine the neighboring atomic environment, are the components of the coordination feature.
Since the interatomic forces corresponding to anionic, cationic, non-ionic, metallic constraints are repulsive, the maximum interatomic distances of the four kinds should be $\infty$.
Therefore, the four kinds of the maximum atomic radii should be defined as
\begin{equation}
R^{\left( \mathrm{A} \right)} = R^{\left( \mathrm{C} \right)} = R^{\left( \mathrm{N_I} \right)} = R^{\left( \mathrm{M} \right)} = \infty.
\end{equation}
The other six kinds of atomic radii are the factors to determine the feasible atomic environments such as the maximum coordination number.

The minimum cationic repulsion radius $r^{\left( \mathrm{C} \right)}$ constrains the feasible connection of coordination polyhedra;
large $r^{\left( \mathrm{C} \right)}$ lengthens the distance between cations resulting in the infeasibility of the edge or face sharing of coordination polyhedra.
This infeasiblity corresponds to large repulsive force between highly charged cations; inorganic structural chemistry considers that highly charged cations prefer the vertex sharing more than the edge or face sharing of coordination polyhedra to reduce the electrostatic repulsion by lengthening the interatomic distance.
Figure \ref{fig:linked_polyhedra} shows the vertex, edge, and face sharings of coordination polyhedra in the quadruple perovskite structure, in which iron, copper, and calcium atoms make the octahedron, square, and regular icosahedron, respectively.
This structure can be reproduced as the optimal solution when all the $r^{\left( \mathrm{C} \right)}$ of iron, copper, and calcium atoms are set to be $C^{\left( \mathrm{Qd} \right)}$, where the discretized feasible atomic radii in this study are defined in Appendix~\ref{sec:discretized_feasible_atomic_radii}.
While the edge and face sharing of two octahedra is infeasible due to the large $r^{\left( \mathrm{C} \right)}$, the egde sharing of square and regular icosahedron and the face sharing of octahedron and regular icosahedron are feasible, because the ionic bond radius of calcium is large.

The rule to assign one out of the types of geometrical constraints to the pair of atoms $i$ and $j \bm{T}$ is mathematically formalized by introducing the creatability function $\Lambda _{\sigma} ^{\left(G \right)} \left( \bm{\Phi}, \varepsilon \right)$ depending on the geometrical constraint $G$ and the spatial order of atoms $\bm{\Phi}$.
Like an activation function in neural networks, this function returns one or more if the geometrical constraint $G$ is creatable.
The inequality constraints using $\Lambda _{\sigma} ^{\left(G \right)} \left( \bm{\Phi}, \varepsilon \right)$ given in the sixth constraint of Eq.~\eqref{eq:equation_of_crystal_structures} gives a choice of creatable types of geometrical constraints, and the suitable one is chosen for the feasibility and/or optimality of the structure.
This study gives a simple definition of $\Lambda _{\sigma} ^{\left(G \right)} \left( \bm{\Phi}, \varepsilon \right)$ as follows:
First, let $\chi _i ^{ \left( \mathcal{E} \right) } \left( \bm{X} \right)$ be the function of
\begin{equation}
\chi _i ^{ \left( \mathcal{E} \right) } \left( \bm{X} \right) =
\begin{cases}
1 & \text{($N_i ^{\left( \mathrm{A} \right)} \in \mathcal{E}$)} \\
0 & \text{($N_i ^{\left( \mathrm{A} \right)} \not\in \mathcal{E}$)}
\end{cases},
\end{equation}
where $N_i ^{\left( \mathrm{A} \right)}$ is the atomic number of atom $i$, $\mathcal{E}$ is the subset of atomic numbers, and $\bm{X}$ is the structure matrix defined by
\begin{equation}
\bm{X} = \left(\bm{x}_1 , \cdots , \bm{x}_n , \bm{\phi}_1, \cdots, \bm{\phi}_n , \bm{t}_1 , \bm{t}_2 , \bm{t}_3 \right).
\end{equation}
with $\bm{\phi}_i$ being the coordination feature vector including atomic number and several kinds of atomic radii.
In this study, the four kinds of subsets $\mathcal{E}$ are defined: Metal atoms $\mathcal{M}$, anions $\mathcal{A}$, pnictogens $\mathcal{P}$, and oxygen $\mathcal{O}$.
Second, let $\Xi _{\sigma} ^{\left( G \right)} \left( \bm{X}, \varepsilon \right)$ be the function defined by
\begin{equation}
\Xi _{\sigma} ^{\left( G \right)} \left( \bm{X}, \varepsilon \right) = \begin{cases}
1 & \text{(satisfy Eq.~\eqref{eq:relaxed_geometrical_constraints})} \\
0 & \text{(otherwise)}
\end{cases}.
\end{equation}
Third, let $\bm{\mathrm{\Phi}}$ be the state tensor of crystal structure defined by
\begin{equation}
\bm{\Phi} = \bm{X} \otimes \bm{n}.
\end{equation}
where $\bm{n}$ is the state vector of geometrical constraints.
Finally, this study defines the creatablity function $\Lambda _{\sigma} ^{\left(G \right)} \left( \bm{\Phi}, \varepsilon \right) \in \left\{ 0,1 \right\}$ as
\begin{equation}
\begin{split}
\Lambda _{\sigma} ^{\left(\mathrm{I} \right)} \left( \bm{\Phi}, \varepsilon \right) &= \left[ \chi_i ^{ \left(\mathcal{M} \right)} \chi _j ^{ \left(\mathcal{A} \right)} + \chi_i ^{ \left(\mathcal{A} \right)} \chi _j ^{ \left(\mathcal{M} \right)} \right] \Xi _{\sigma} ^{\left(\mathcal{I} \right)} \left( \bm{X}, \varepsilon \right), \\
\Lambda _{\sigma} ^{\left(\mathrm{N_I} \right)} \left( \bm{\Phi}, \varepsilon \right) &= \chi_i ^{ \left(\mathcal{M} \right)} \chi _j ^{ \left(\mathcal{A} \right)} + \chi_i ^{ \left(\mathcal{A} \right)} \chi _j ^{ \left(\mathcal{M} \right)}, \\
\Lambda _{\sigma} ^{\left(\mathrm{A} \right)} \left( \bm{\Phi}, \varepsilon \right) &= \chi_i ^{ \left(\mathcal{A} \right)} \chi _j ^{ \left(\mathcal{A} \right)}, \\
\Lambda _{\sigma} ^{\left(\mathrm{C} \right)} \left( \bm{\Phi}, \varepsilon \right) &= \begin{cases}
\chi_i ^{ \left(\mathcal{M} \right)} \chi _j ^{ \left(\mathcal{M} \right)} & \text{($n_{\sigma} ^{\left( \mathrm{common} \right)} > 0$)} \\
0 & \text{($n_{\sigma} ^{\left( \mathrm{common} \right)} = 0$)}
\end{cases}, \\
\Lambda _{\sigma} ^{\left(\mathrm{N_M} \right)} \left( \bm{\Phi}, \varepsilon \right) &= \begin{cases}
\chi_i ^{ \left(\mathcal{M} \right)} \chi _j ^{ \left(\mathcal{M} \right)} & \text{($n_{\sigma} ^{\left( \mathrm{common} \right)} = 0$)} \\
0 & \text{($n_{\sigma} ^{\left( \mathrm{common} \right)} > 0$)}
\end{cases},
\end{split} \label{eq:constraint_creatability}
\end{equation}
where $n_{\sigma} ^{\left( \mathrm{common} \right)}$ is the number of common bridging anions between metals $i$ and $j \bm{T}$ that can be formalized as
\begin{equation}
n_{\sigma} ^{\left( \mathrm{common} \right)} = \chi_i ^{ \left(\mathcal{M} \right)} \chi_j ^{ \left(\mathcal{M} \right)} \sum_{k, \bm{T}^{\prime}} \chi_k ^{ \left(\mathcal{A} \right)} n_{i k \bm{T}^{\prime}} ^{\left(\mathrm{I} \right)} \, n_{j k \left(\bm{T}^{\prime} - \bm{T} \right)} ^{\left(\mathrm{I} \right)}.
\end{equation}
Note that $n_{\sigma} ^{\left( \mathrm{common} \right)}$ corresponds to the connection of coordination polyhedra as
\begin{equation}
\begin{cases}
n_{\sigma} ^{\left( \mathrm{common} \right)} = 1 & \text{(Vertex sharing of polyhedra)} \\
n_{\sigma} ^{\left( \mathrm{common} \right)} = 2 & \text{(Edge sharing of polyhedra)} \\
n_{\sigma} ^{\left( \mathrm{common} \right)} \ge 3 & \text{(Face sharing of polyhedra)}
\end{cases}.
\end{equation}
The creatability function in this study returns one if the geometrical constraint is creatable but zero otherwise.
Both the ionic and non-ionic bonds can be formed between an anion and a cation, but the ionic bond has the additional condition that the interatomic distance has to be feasible if it is formed.
Note that if $R^{\left( \mathrm{I} \right)} < r^{\left( \mathrm{N_I} \right)}$, the two kinds of the constraints satisfy
\begin{equation}
\Lambda _{\sigma} ^{\left(\mathrm{I} \right)} \left( \bm{\Phi}, 0 \right) \cdot \Lambda _{\sigma} ^{\left(\mathrm{N_I} \right)} \left( \bm{\Phi}, 0 \right) = 0.
\end{equation}
In this study, $r^{\left( \mathrm{N_I} \right)}$ is defined as
\begin{equation}
r^{\left( \mathrm{N_I} \right)} = \gamma ^{\left( \mathrm{N_I} \right)} R^{\left( \mathrm{I} \right)},
\end{equation}
where $\gamma ^{\left( \mathrm{N_I} \right)}$ is a coefficient.
The constant is set to be $1.2$ for the random structure search but $1.0$ for the analysis of oxide crystals registered in ICSD.
On the other hand, while anionic constraints can always be formed between two anions, cationic constraints cannot be formed unless common bridging anions exist and metallic constraints vise versa.
Finally, the sum of the number of created geometrical constraints assigned to the pair of the atoms $i$ and $j \bm{T}$ must be one:
\begin{equation}
\sum_{G} n_{\sigma} ^{\left(G \right)} = 1,
\end{equation}
where $n_{\sigma} ^{\left(G \right)} \in \left\{ 0,1 \right\}$ is the number of created geometrical constraints $G$ assigned to the atoms $i$ and $j \bm{T}$.

The mathematical optimization problem searching structural prototypes of ionic compounds is formalized as follows:
First, the structure minimizes the volume of unit cell $\Omega$.
Second, all the interatomic distances must be feasible.
Third to sixth, one out of the types of the geometrical constraints is assigned to every pair of atoms, and the minimum and maximum distances are the sums of the minimum and maximum atomic radii corresponding to the assigned geometrical constraints, respectively.
Seventh, a choice of the creatable types of geometrical constraints for every pair of atoms is given.
Accordingly, the optimization problem for ionic crystals is given by
\begin{equation}
\begin{split}
\text{minimize} \quad & \Omega \\
\text{subject to} \quad & \left( 1 - \varepsilon \right) d_{\sigma} \le x_{\sigma} \le \left( 1 + \varepsilon \right) D_{\sigma} \\
& d_{\sigma} = \max_{G} \left[ n_{\sigma} ^{\left(G \right)} \left( r_i ^{\left(G \right)} + r_j ^{\left(G \right)} \right) \right] \\
& D_{\sigma} = \min_{G} \left[ n_{\sigma} ^{\left(G \right)} \left( R_i ^{\left(G \right)} + R_j ^{\left(G \right)} \right) \right] \\
& \sum_{G} n_{\sigma} ^{\left(G \right)} = 1 \\
& n_{\sigma} ^{\left(G \right)} \in \left\{ 0, 1 \right\} \\
& n_{\sigma} ^{\left(G \right)} \le \Lambda _{\sigma} ^{\left(G \right)} \left( \bm{\Phi}, \varepsilon \right)
\end{split}
\label{eq:equation_of_crystal_structures}
\end{equation}
Note that if $\Lambda _{\sigma} ^{\left(\mathrm{I} \right)} \left( \bm{\Phi}, \varepsilon \right) = 1$ with $\varepsilon = 0$, the ionic bond must be formed if $R^{\left( \mathrm{I} \right)} < r^{\left( \mathrm{N_I} \right)}$.
In that case, the geometrical constraints are deterministically assigned to all the pairs of atoms.

In this study, the coordination numbers of every cation are fixed to accelerate the discovery of optimal structures by limiting the feasible coordination polyhedra, while anions have no constraint on the coordination numbers.
Let $\bm{N}^{\left( \mathrm{I} \right)}$ be the vector of the fixed coordination numbers depending on the subsets of atomic numbers, for example, if a cation is surrounded by four oxide ions and four pnictogen ions, the coordination vector is given by $\left(N_{\mathcal{O}} ^{\left( \mathrm{I} \right)}, N_{\mathcal{P}} ^{\left( \mathrm{I} \right)} \right) = \left( 4,4 \right)$.
Since the creatability functions $\Lambda _{\sigma} ^{\left(\mathrm{I} \right)} \left( \bm{\Phi}, \varepsilon \right)$ given in Eq.~\eqref{eq:constraint_creatability} do not use the information on the coordination numbers, this study marks up the additional constraints in Eq.~\eqref{eq:equation_of_crystal_structures} given by
\begin{equation}
\sum_{j, \bm{T}} \chi_j ^{\left( \mathcal{E} \right)} n_{\sigma} ^{\left(\mathrm{I} \right)} = N_{i \mathcal{E}} ^{\left(\mathrm{I} \right)}.
\label{eq:coordination_number_constraint}
\end{equation}
This equation constrains the coordination numbers depending on the subsets of atomic numbers.
If an atom does not satisfy the constraints, the structure is regarded as an infeasible solution.

\section{Methods}
\label{sec:methods}

\begin{figure*}
\centering
\includegraphics[width=2.0\columnwidth]{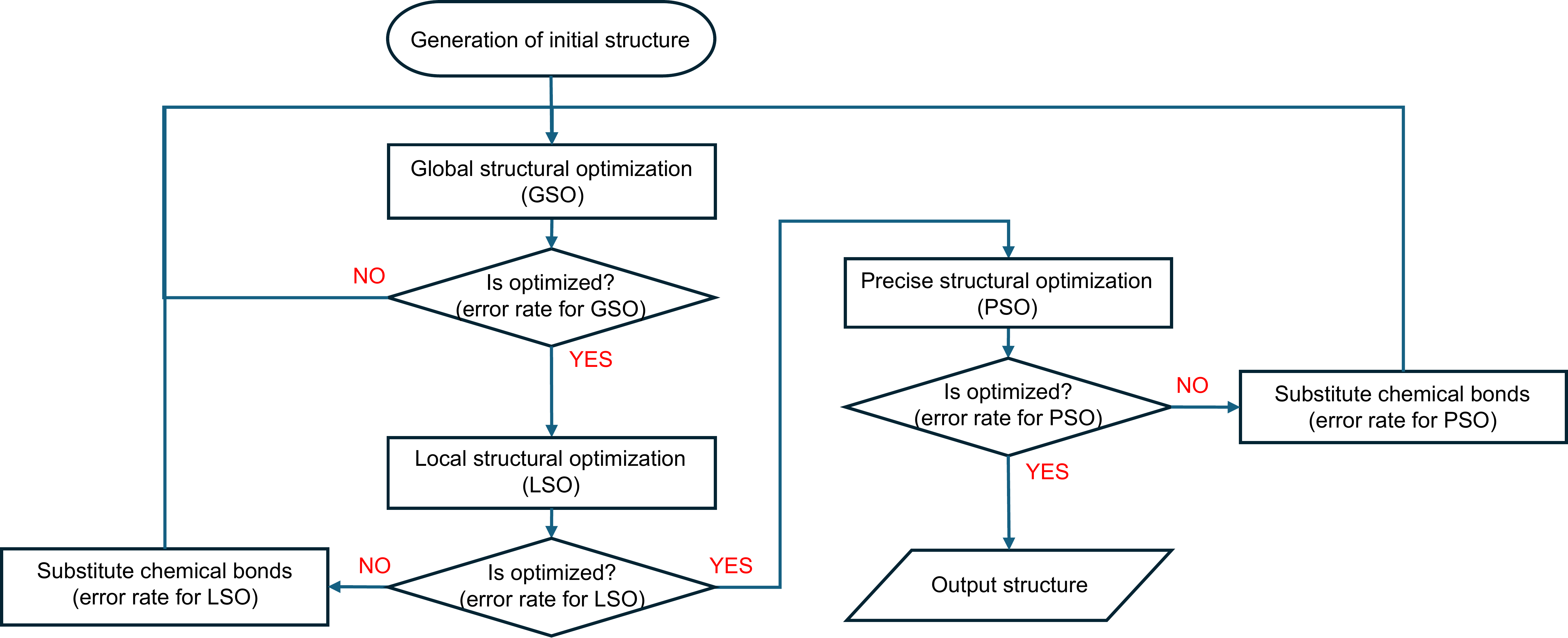}
\caption{
The flow chart of the algorithm to find the optimal solutions of the problem given in Eq.~\eqref{eq:equation_of_crystal_structures} from random sampling.
In the global optimization, $\bm{X}$ and $\bm{n}$ are alternately optimized, while in the local and precise optimization, only $\bm{X}$ is optimized to identify whether all the geometrical constraints can be feasible.
}
\label{fig:structural_optimization_flowchart}
\end{figure*}

The flowchart to solve the problem of Eq.~\eqref{eq:equation_of_crystal_structures} with Eq.~\eqref{eq:coordination_number_constraint} from random sampling is shown in Fig.~\ref{fig:structural_optimization_flowchart}.
How to generate initial structures are discussed in Appendix~\ref{sec:initial_structure_generation}.
The algorithm is composed of the three parts of the global, local, and precise optimization, which have larger error rate $\varepsilon$ from left to right.
If a structure is optimal under an error rate, the structure is judged whether it can be an optimal solution under smaller error rate by optimizing with smaller displacement sizes of atoms and crystal lattice.
The global optimization is aimed at finding different network of ionic bonds by large deformation of the structure, while the local and precise optimization is aimed at judging whether all the geometrical constraints can be feasible.
Therefore, $\bm{X}$ and $\bm{n}$ are alternately optimized in the global optimization to escape from infeasible region with large displacement size, while the optimization on $\bm{n}$ is only applied at the beginning in the local and precise optimization.
Note that the optimization of $\bm{X}$ with fixed $\bm{n}$ corresponds to the structural optimization whose algorithm is summarized in Appendix~\ref{sec:structural_optimization_algorithm}, while the optimization of $\bm{n}$ with fixed $\bm{X}$ corresponds to the replacement of the geometrical constraints.

In many cases, it is difficult for a structure to escape from the infeasible region by only the alternative optimization of $\bm{X}$ and $\bm{n}$.
Therefore, if a structure is regarded as an infeasible structure after the local or precise optimization, the ionic bonds are randomly substituted as follows:
First, one cation owing infeasible ionic bonds is randomly chosen.
Note that if a cation does not satisfy the implicit constraints on the feasible connection of coordination polyhedra, some ionic bonds are stretched after the structural optimization, because the repulsive force is dominant compared to the attractive force and the pressure as discussed in Appendix~\ref{sec:structural_optimization_algorithm}.
Second, the cation randomly choose one of owing non-ionic bonds.
Third, all the owing ionic bonds are converted into non-ionic bonds.
Finally, the chosen non-ionic bond is converted into ionic bond.
This operation enables the infeasible cation to move into a neighboring site of the packing of oxide ions.

How to replace the types of geometrical constraints is simple:
First, every cation creates as many creatable ionic bonds as possible under the constraint on the coordination numbers.
Second, if a pair of an anion and a cation does not have an ionic bond, a non-ionic bond is formed between the pair.
Third, cationic constraints are formed if the two cations have common bridging anions.
Fourth, if a pair of two cations does not have a cationic constraint, a metallic constraint is formed between the pair.
Fifth, anionic constraints are formed between every pair of anions.
This operation can be mathematically formalized as
\begin{equation}
\begin{split}
\text{maximize} \quad & \sum_i \sum_{j \bm{T}} \chi_i ^{\left( \mathcal{M} \right)} n_{\sigma} ^{\left( \mathrm{I} \right)} \\
& \sum_{G} n_{\sigma} ^{\left(G \right)} = 1 \\
& n_{\sigma} ^{\left(G \right)} \in \left\{ 0, 1 \right\} \\
& n_{\sigma} ^{\left(G \right)} \le \Lambda _{\sigma} ^{\left(G \right)} \left( \bm{X}, \varepsilon \right) \\
& \sum_{j, \bm{T}} \chi_j ^{\left( \mathcal{E} \right)} n_{\sigma} ^{\left(\mathrm{I} \right)} \le N_{i \mathcal{E}} ^{\left(\mathrm{I} \right)}
\end{split}.
\label{eq:maximization_problem_of_ionic_bonds}
\end{equation}
Note that a cation does not necessarily satisfy the coordination numbers, because an ionic bond is getting infeasible if the interatomic distance between the cation and anion is larger.
Besides, in this study, if a cation has more neighboring anions than the coordination number, the cation chooses the fixed number of anions in the order of the interatomic distances.

\section{Result and discussion}
\label{sec:result_and_discussion}

\begin{figure}
\centering
\includegraphics[width=1.0\columnwidth]{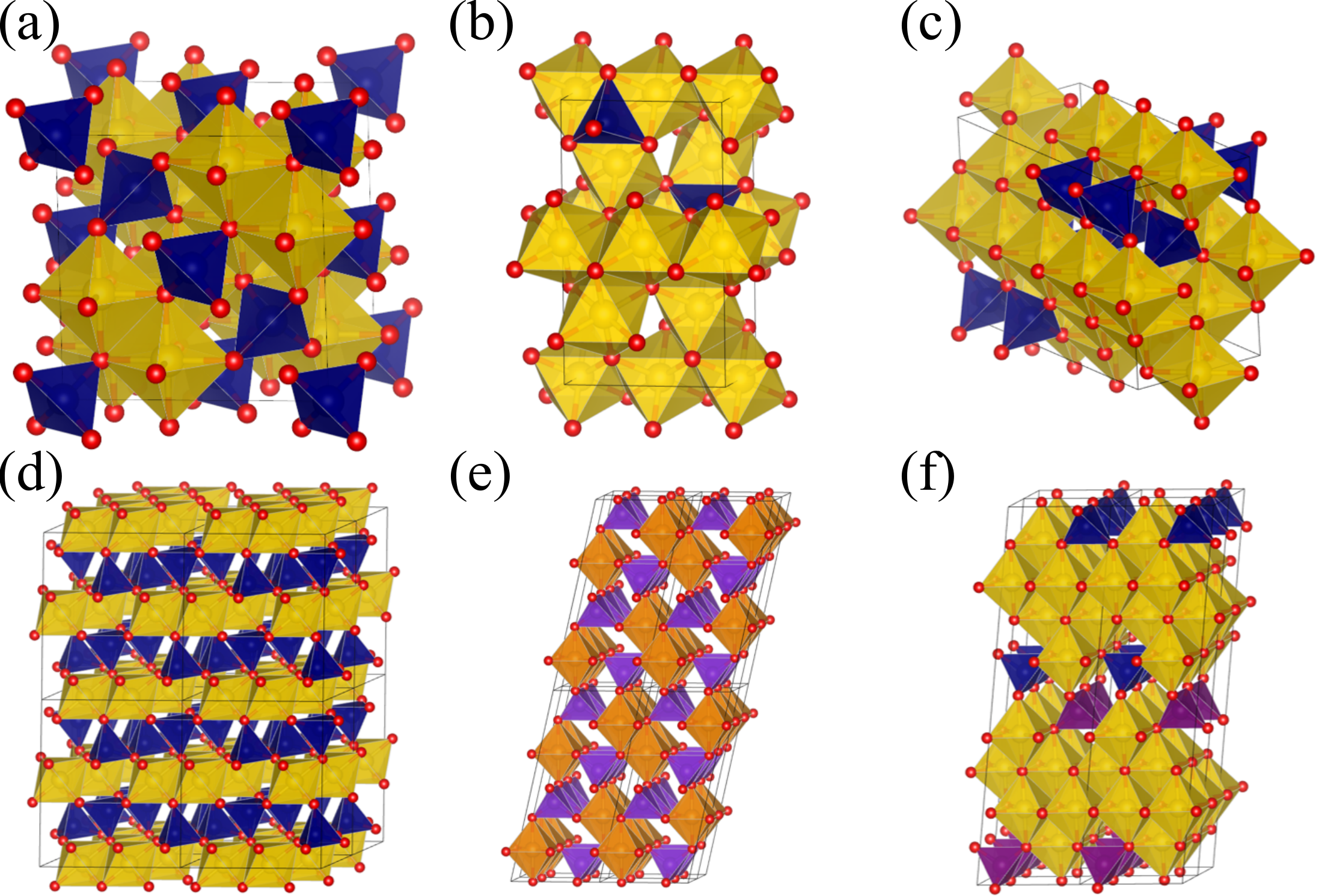}
\caption{
Silocon atoms are shown as darkblue balls.
A silicon atom makes a tetrahedron, while an iron atom makes an octahedron.
(a) The spinel structure of \ch{Si4Fe8O16}.
(b) The \ch{Mg2SiO4} structure of \ch{Si4Fe8O16}.
(c) The $\beta$-\ch{Mg2SiO4} structure of \ch{Si4Fe8O16}.
(d) The \ch{CaMgSi2O6} structure of \ch{Si4Fe4O12}.
(f) The \ch{Ga2O3} structure of \ch{Ga4^{[4$t$]}Ga4^{[6o]}O12}. The blueviolet and darkorange balls correspond to galium atoms whose coordination number is four and six, respectively. The former makes a tetrahedron, while the latter makes a octahedron.
(f) The \ch{SiFe7O10} structure of \ch{Si2Fe14O20}. The purple balls correspond to iron atoms whose coordination number is four.
}
\label{fig:iron_oxides}
\end{figure}

\begin{figure}
\centering
\includegraphics[width=1.0\columnwidth]{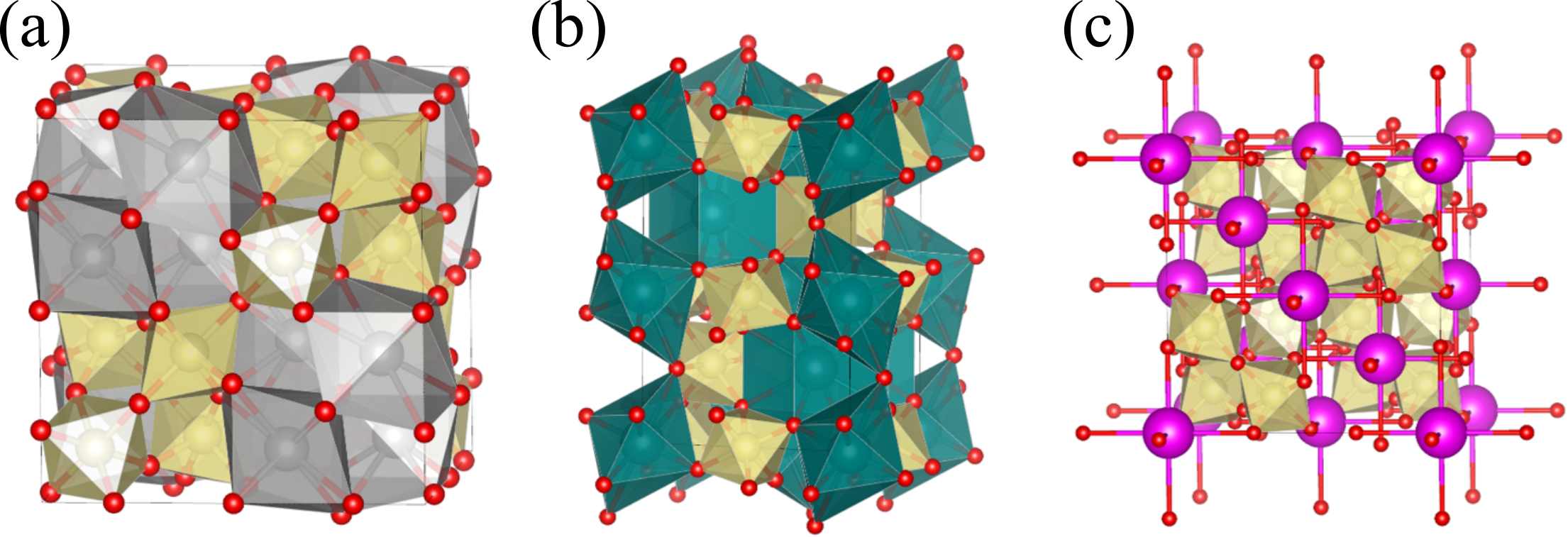}
\caption{
(a) The $\alpha$-pyrochlore structure of \ch{Hg4Os4O14}. Mercury and osmium atoms are shown as darkgray and khaki balls, respectively. A mercury atom makes a distorted hexagonal bipyramid, while an osmium atom makes an octahedron.
(b) The weberite structure of \ch{Ca4Os4O14}. A calcium atom makes a hexagonal bipyramid, while an osmium atom makes an octahedron.
(c) The $\beta$-pyrochlore structure of \ch{Rb2Os4O12}. Rubidium atoms are shown as magenta balls, respectively. A rubidium atom makes a large octahedron, while an osmium atom makes a closed-packed octahedron.
}
\label{fig:osmium_oxides}
\end{figure}

\begin{figure}
\centering
\includegraphics[width=1.0\columnwidth]{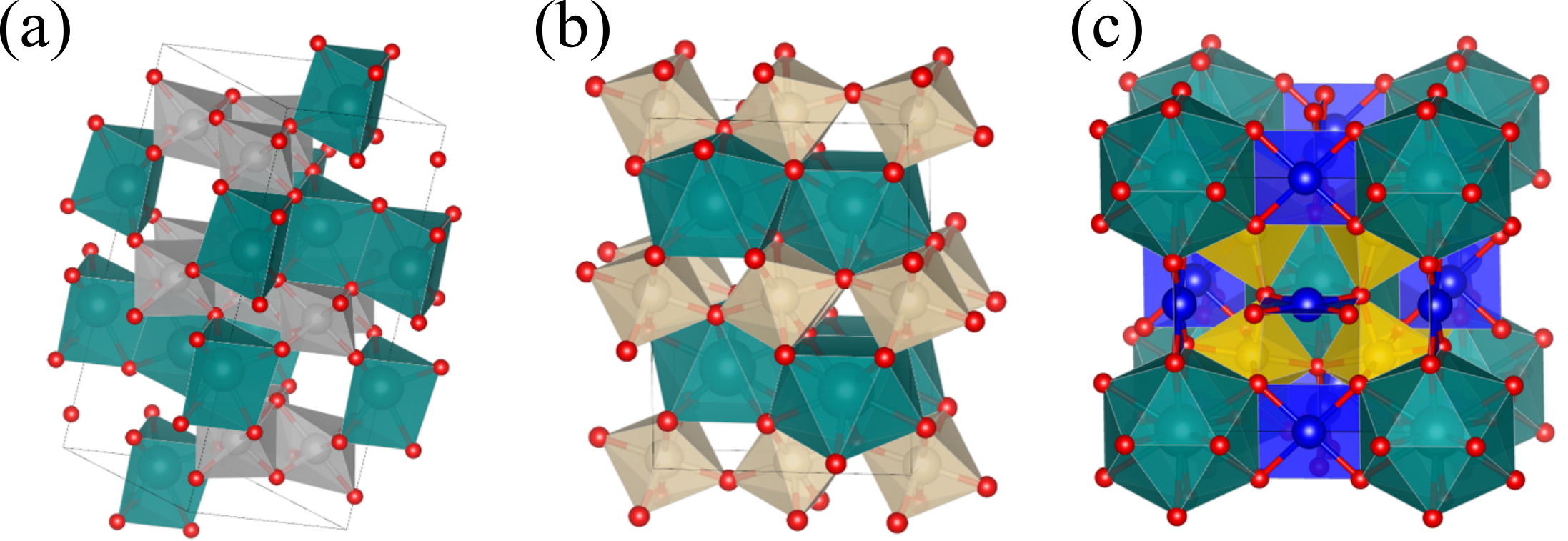}
\caption{
(a) The \ch{Na2Mn3O8} structure of \ch{Ca2Pt3O8}. Platinum atoms are shown as silver balls. A platinum atom makes an octahedron, while a calcium atom makes a trigonal prism.
(b) The pyroxene structure of \ch{Ca4Ir4O12}. Iridium atoms are shown as wheat balls. An iridium atom makes an octahedron, while a calcium atom makes a distorted square antiprism.
(c) The quadruple perovskite structure of \ch{CaCu3Fe4O12}. A copper atom makes a square, an iron atom makes an octahedron, and a calcium atom makes a regular icosahedron.
}
\label{fig:calcium_oxides}
\end{figure}

\begin{figure}
\centering
\includegraphics[width=1.0\columnwidth]{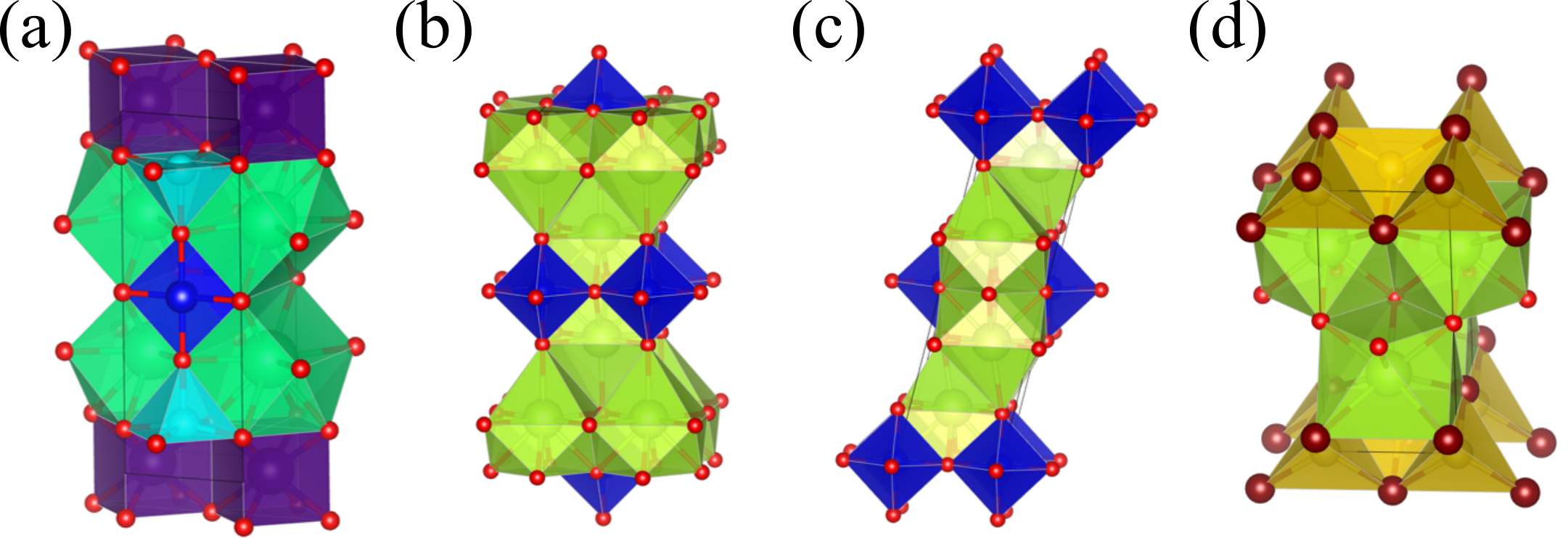}
\caption{
(a) The \ch{YBa2Cu3O_{7-$x$}} structure of \ch{YBa2Cu3O7}. Yttrium and barium atoms are shown as indigo and springgreen balls, respectively. The former makes the cubic while the latter is surrounded by ten oxide ions. The aqua balls correspond to copper atoms which makes the square pyramid, while the blue balls correspond to copper atoms which makes the square.
(b) The \ch{K2NiF4} structure of \ch{La4Cu2O8}. Lanthanum atoms are shown as greenyellow balls. A lanthanum atom makes a capped square antiprism, while a copper atom makes an octahedron.
(c) A distorted \ch{K2NiF4} structure with $C2/m$ symmetry. Lanthanum atoms are shown as greenyellow balls. A lanthanum atom makes a distorted capped square antiprism, while a copper atom makes an octahedron.
(d) The \ch{ZrCuSiAs} structure of \ch{La2Fe2As2O2}. Arsenic atoms are shown as garkred balls. A lanthanum atom makes a square antiprism, while an iron atom makes a tetrahedron.
}
\label{fig:oxide_superconductors}
\end{figure}

\begin{figure*}
\centering
\includegraphics[width=1.8\columnwidth]{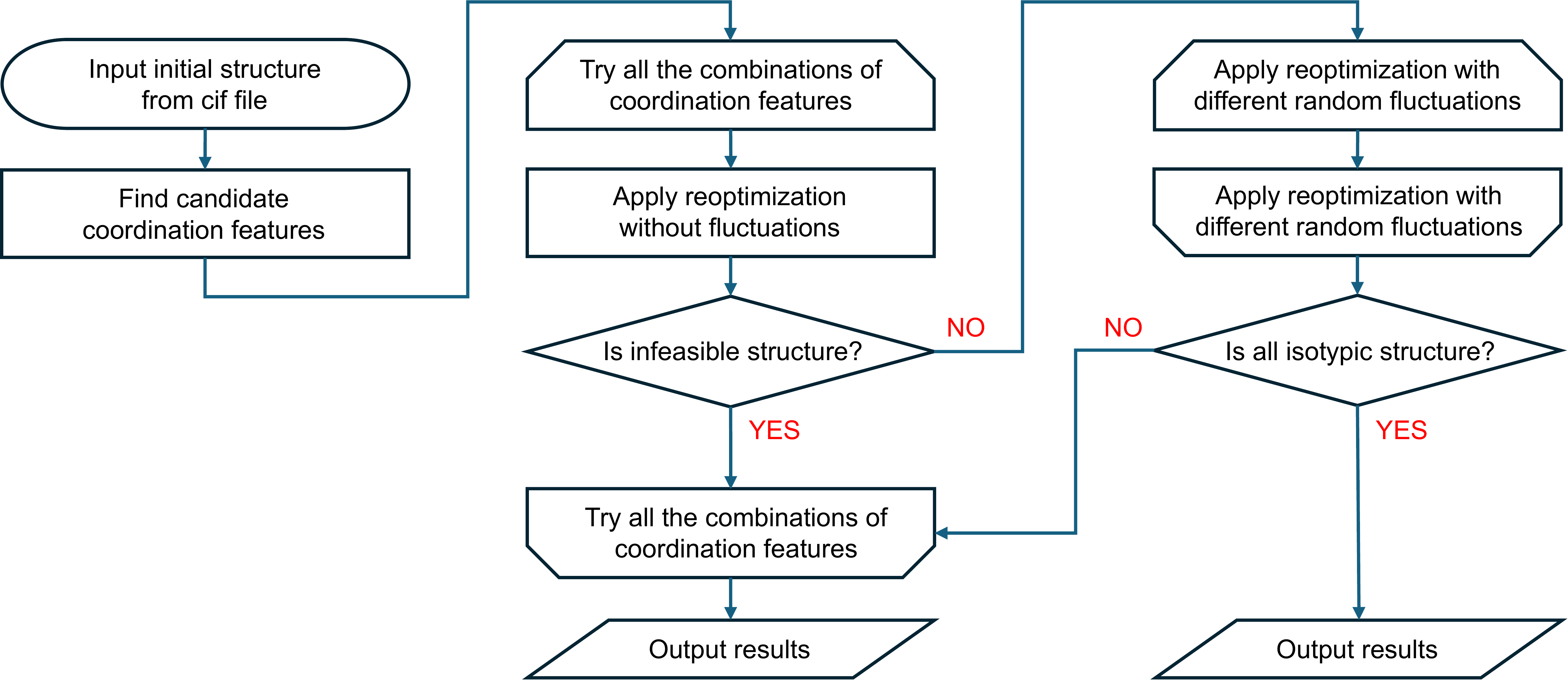}
\caption{
The flow chart of the scheme to find the best degree of structural similarity between the experimental and reoptimized structures.
All the promising combination of coordination features are analyzed whether the reoptimized structure can be an optimal solution with structural similarity until the isotypic optimal solution is found.
}
\label{fig:matching degree_check_flowchart}
\end{figure*}

\begin{table*}
\caption{
The chemical compositions to which the mathematial optimization problem is applied.
The coordination feature, the number of randomly generated structures, and the discovered optimal solutions are also listed.
A coordination features is written as $\left(N^{\left( \mathrm{A} \right)}, \mathcal{E}, R^{\left( \mathrm{I} \right)}, r^{\left( \mathrm{I} \right)}, R^{\left( \mathrm{R} \right)}, \bm{N}_I \right)$ with $\mathrm{R} \in \left\{ \mathrm{A}, \mathrm{C}\right\}$, where the atomic number $N^{\left( \mathrm{A} \right)}$ is expressed by element symbol for a straightforward manner.
Oxygen has the coordination feature given by $\left(\ch{O}, \mathcal{O}, C^{\left( \mathrm{O} \right)}, C^{\left( \mathrm{O} \right)}, C^{\left( \mathrm{O} \right)}, \mathcal{M}_{\ast} \right)$, where oxygen can create ionic bonds with metal atoms freely as expressed by $\bm{N}_I = \left(\mathcal{M}_{\ast} \right)$.
}
\label{table:target_chemical_compositions}
\begin{ruledtabular}
\begin{tabular}{cccc}
Chemical composition & Coordination features & Number of generated structures & Discovered optimal solutions \\ \hline
\ch{Si4Fe8O16} & 
\begin{tabular}{l}
$\left(\ch{Fe}, \mathcal{M}, C^{\left( \mathrm{H} \right)}, C^{\left( \mathrm{H} \right)}, C^{\left( \mathrm{O} \right)}, \mathcal{O}_6 \right)$ \\
$\left(\ch{Si}, \mathcal{M}, C^{\left( \mathrm{Q} \right)}, C^{\left( \mathrm{Q} \right)}, C^{\left( \mathrm{Pd} \right)}, \mathcal{O}_4 \right)$
\end{tabular}
& $5 \times 10^{6}$ & spinel structure \\ \hline
\ch{Si4Fe8O16} & 
\begin{tabular}{l}
$\left(\ch{Fe}, \mathcal{M}, C^{\left( \mathrm{H} \right)}, C^{\left( \mathrm{H} \right)}, C^{\left( \mathrm{O} \right)}, \mathcal{O}_6 \right)$ \\
$\left(\ch{Si}, \mathcal{M}, C^{\left( \mathrm{Q} \right)}, C^{\left( \mathrm{Q} \right)}, C^{\left( \mathrm{O} \right)}, \mathcal{O}_4 \right)$
\end{tabular}
& $1 \times 10^{7}$ &
\begin{tabular}{l}
spinel structure \\
$\beta$-\ch{Mg2SiO4} structure \\
other $207$ structures
\end{tabular} \\ \hline
\ch{Si4Fe8O16} & 
\begin{tabular}{l}
$\left(\ch{Fe}, \mathcal{M}, C^{\left( \mathrm{H} \right)}, C^{\left( \mathrm{H} \right)}, C^{\left( \mathrm{O} \right)}, \mathcal{O}_6 \right)$ \\
$\left(\ch{Si}, \mathcal{M}, C^{\left( \mathrm{Q} \right)}, C^{\left( \mathrm{Q} \right)}, C^{\left( \mathrm{E} \right)}, \mathcal{O}_4 \right)$
\end{tabular}
& $1 \times 10^{6}$ &
\begin{tabular}{l}
spinel structure \\
$\beta$-\ch{Mg2SiO4} structure \\
\ch{Mg2SiO4} structure \\
other $5590$ structures
\end{tabular} \\ \hline
\ch{Si4Fe4O12} & 
\begin{tabular}{l}
$\left(\ch{Fe}, \mathcal{M}, C^{\left( \mathrm{H} \right)}, C^{\left( \mathrm{H} \right)}, C^{\left( \mathrm{O} \right)}, \mathcal{O}_6 \right)$ \\
$\left(\ch{Si}, \mathcal{M}, C^{\left( \mathrm{Q} \right)}, C^{\left( \mathrm{Q} \right)}, C^{\left( \mathrm{O} \right)}, \mathcal{O}_4 \right)$
\end{tabular}
& $5 \times 10^{6}$ &
\begin{tabular}{l}
\ch{CaMgSi2O6} structure \\
\ch{Ga2O3} structure \\
other $602$ structures
\end{tabular} \\ \hline
\ch{Si2Fe14O20} & 
\begin{tabular}{l}
$\left(\mathrm{Fe}, \mathcal{M}, C^{\left( \mathrm{H} \right)}, C^{\left( \mathrm{H} \right)}, C^{\left( \mathrm{O} \right)}, \mathcal{O}_6 \right)$ \\
$\left(\mathrm{Si}, \mathcal{M}, C^{\left( \mathrm{Q} \right)}, C^{\left( \mathrm{Q} \right)}, C^{\left( \mathrm{O} \right)}, \mathcal{O}_4 \right)$
\end{tabular}
& $1 \times 10^{6}$ &
\begin{tabular}{l}
\ch{SiFe7O10} structure \\
other $66$ structures
\end{tabular} \\ \hline
\ch{Hg4Os4O14} & 
\begin{tabular}{l}
$\left(\mathrm{Hg}, \mathcal{M}, C^{\left( \mathrm{Wd} \right)}, C^{\left( \mathrm{S} \right)}, C^{\left( \mathrm{Hd} \right)}, \mathcal{O}_8 \right)$ \\
$\left(\mathrm{Os}, \mathcal{M}, C^{\left( \mathrm{H} \right)}, C^{\left( \mathrm{H} \right)}, C^{\left( \mathrm{Hd} \right)}, \mathcal{O}_6 \right)$
\end{tabular}
& $5 \times 10^{6}$ & $\alpha$-pyrochlore structure \\ \hline
\ch{Ca4Os4O14} & 
\begin{tabular}{l}
$\left(\mathrm{Ca}, \mathcal{M}, C^{\left( \mathrm{Td} \right)}, C^{\left( \mathrm{S} \right)}, C^{\left( \mathrm{Hd} \right)}, \mathcal{O}_8 \right)$ \\
$\left(\mathrm{Os}, \mathcal{M}, C^{\left( \mathrm{H} \right)}, C^{\left( \mathrm{H} \right)}, C^{\left( \mathrm{Hd} \right)}, \mathcal{O}_6 \right)$
\end{tabular}
& $5 \times 10^{6}$ &
\begin{tabular}{l}
$\alpha$-pyrochlore structure \\
weberite structure \\
other $6$ structures
\end{tabular} \\ \hline
\ch{Rb2Os4O12} & 
\begin{tabular}{l}
$\left(\mathrm{Rb}, \mathcal{M}, C^{\left( \mathrm{Qv} \right)}, C^{\left( \mathrm{Sd} \right)}, C^{\left( \mathrm{V} \right)}, \mathcal{O}_{18} \right)$ \\
$\left(\mathrm{Os}, \mathcal{M}, C^{\left( \mathrm{H} \right)}, C^{\left( \mathrm{H} \right)}, C^{\left( \mathrm{Hd} \right)}, \mathcal{O}_6 \right)$
\end{tabular}
& $5 \times 10^{6}$ & $\beta$-pyrochlore structure \\ \hline
\ch{Ca2Pt3O8} & 
\begin{tabular}{l}
$\left(\mathrm{Ca}, \mathcal{M}, C^{\left( \mathrm{E} \right)}, C^{\left( \mathrm{E} \right)}, C^{\left( \mathrm{Pd} \right)}, \mathcal{O}_6 \right)$ \\
$\left(\mathrm{Pt}, \mathcal{M}, C^{\left( \mathrm{S} \right)}, C^{\left( \mathrm{H} \right)}, C^{\left( \mathrm{Td} \right)}, \mathcal{O}_6 \right)$
\end{tabular}
& $1 \times 10^{6}$ &
\begin{tabular}{l}
\ch{Na2Mn3O8} structure \\
other $575$ structures
\end{tabular} \\ \hline
\ch{Ca4Ir4O12} & 
\begin{tabular}{l}
$\left(\mathrm{Ca}, \mathcal{M}, C^{\left( \mathrm{N} \right)}, C^{\left( \mathrm{E} \right)}, C^{\left( \mathrm{Td} \right)}, \mathcal{O}_8 \right)$ \\
$\left(\mathrm{Ir}, \mathcal{M}, C^{\left( \mathrm{H} \right)}, C^{\left( \mathrm{Q} \right)}, C^{\left( \mathrm{Td} \right)}, \mathcal{O}_6 \right)$
\end{tabular}
& $5 \times 10^{6}$ & pyroxene structure \\ \hline
\ch{CaCu3Fe4O12} & 
\begin{tabular}{l}
$\left(\mathrm{Ca}, \mathcal{M}, C^{\left( \mathrm{Wd} \right)}, C^{\left( \mathrm{Wd} \right)}, C^{\left( \mathrm{Qd} \right)}, \mathcal{O}_{12} \right)$ \\
$\left(\mathrm{Cu}, \mathcal{M}, C^{\left( \mathrm{H} \right)}, C^{\left( \mathrm{H} \right)}, C^{\left( \mathrm{Qd} \right)}, \mathcal{O}_4 \right)$ \\
$\left(\mathrm{Fe}, \mathcal{M}, C^{\left( \mathrm{H} \right)}, C^{\left( \mathrm{H} \right)}, C^{\left( \mathrm{Qd} \right)}, \mathcal{O}_6 \right)$
\end{tabular}
& $5 \times 10^{6}$ & quadruple perovskite structure \\ \hline
\ch{YBa2Cu^{[4$l$]}Cu2^{[5$y$]}O7} & 
\begin{tabular}{l}
$\left(\mathrm{Y}, \mathcal{M}, C^{\left( \mathrm{N} \right)}, C^{\left( \mathrm{E} \right)}, C^{\left( \mathrm{Qd} \right)}, \mathcal{O}_{8} \right)$ \\
$\left(\mathrm{Ba}, \mathcal{M}, C^{\left( \mathrm{Td} \right)}, C^{\left( \mathrm{O} \right)}, C^{\left( \mathrm{Qd} \right)}, \mathcal{O}_{10} \right)$ \\
$\left(\mathrm{Cu}, \mathcal{M}, C^{\left( \mathrm{H} \right)}, C^{\left( \mathrm{H} \right)}, C^{\left( \mathrm{Pd} \right)}, \mathcal{O}_4 \right)$ \\
$\left(\mathrm{Cu}, \mathcal{M}, C^{\left( \mathrm{H} \right)}, C^{\left( \mathrm{H} \right)}, C^{\left( \mathrm{Qd} \right)}, \mathcal{O}_5 \right)$
\end{tabular}
& $1 \times 10^{6}$ & \ch{YBa2Cu3O_{7-$x$}} structure \\ \hline
\ch{Y2Ba4Cu2^{[4$l$]}Cu4^{[5$y$]}O14} & 
\begin{tabular}{l}
$\left(\mathrm{Y}, \mathcal{M}, C^{\left( \mathrm{N} \right)}, C^{\left( \mathrm{E} \right)}, C^{\left( \mathrm{Qd} \right)}, \mathcal{O}_{8} \right)$ \\
$\left(\mathrm{Ba}, \mathcal{M}, C^{\left( \mathrm{Td} \right)}, C^{\left( \mathrm{O} \right)}, C^{\left( \mathrm{Qd} \right)}, \mathcal{O}_{10} \right)$ \\
$\left(\mathrm{Cu}, \mathcal{M}, C^{\left( \mathrm{H} \right)}, C^{\left( \mathrm{H} \right)}, C^{\left( \mathrm{Pd} \right)}, \mathcal{O}_4 \right)$ \\
$\left(\mathrm{Cu}, \mathcal{M}, C^{\left( \mathrm{H} \right)}, C^{\left( \mathrm{H} \right)}, C^{\left( \mathrm{Qd} \right)}, \mathcal{O}_5 \right)$
\end{tabular}
& $5 \times 10^{6}$ &
\begin{tabular}{l}
\ch{YBa2Cu3O_{7-$x$}} structure \\
other $5$ structure
\end{tabular} \\ \hline
\ch{La4Cu2O8} & 
\begin{tabular}{l}
$\left(\mathrm{Cu}, \mathcal{M}, C^{\left( \mathrm{H} \right)}, C^{\left( \mathrm{H} \right)}, C^{\left( \mathrm{O} \right)}, \mathcal{O}_{6} \right)$ \\
$\left(\mathrm{La}, \mathcal{M}, C^{\left( \mathrm{O} \right)}, C^{\left( \mathrm{O} \right)}, C^{\left( \mathrm{Sd} \right)}, \mathcal{O}_{9} \right)$
\end{tabular}
& $5 \times 10^{5}$ &
\begin{tabular}{l}
\ch{K2NiF4} structure \\
Distorted \ch{K2NiF4} structure
\end{tabular} \\ \hline
\ch{La2Fe2As2O2} & 
\begin{tabular}{l}
$\left(\mathrm{La}, \mathcal{M}, C^{\left( \mathrm{O} \right)}, C^{\left( \mathrm{O} \right)}, C^{\left( \mathrm{Nd} \right)}, \mathcal{O}_{4} \mathcal{P}_4 \right)$ \\
$\left(\mathrm{Fe}, \mathcal{M}, C^{\left( \mathrm{H} \right)}, C^{\left( \mathrm{H} \right)}, C^{\left( \mathrm{O} \right)}, \mathcal{P}_{4} \right)$ \\
$\left(\mathrm{As}, \mathcal{P}, C^{\left( \mathrm{Ed} \right)}, C^{\left( \mathrm{Ed} \right)}, C^{\left( \mathrm{Ed} \right)}, \mathcal{M}_{\ast} \right)$
\end{tabular}
& $1 \times 10^{6}$ & \ch{ZrCuSiAs} structure \\
\end{tabular}
\end{ruledtabular}
\end{table*}

\begin{table*}
\caption{
The definitions of the degree of structural similarity between two structures.
In addition to the space group type and the Wyckoff sequence, which are utilized in the previous studies for identifying inorganic structural types~\cite{https://doi.org/10.1107/S0108767389008834, Allmann:sh0188}, the geometrical constraints are also utilized to identify the coordination numbers and the connections of coordination polyhedra.
The structure fingerprint (SF) and the simple structure fingerprint (SSF) are defined in Appendix~\ref{sec:structure_fingerprint}.
}
\label{table:definitions_of_matching degree}
\begin{ruledtabular}
\begin{tabular}{cc}
Coincidence & Definition \\ \hline
Isotypic & Optimal solution has the same SF. \\
Homeotypic & Optimal solution has the different space group type or site symmetries, but the same SSF. \\
Isopointal & Optimal solution has the same space group type and site symmetries, but different SSF. \\
Optimal & Optimal solution has a different space group type and SSF. \\
Infeasible & An infeasible solution. \\
\end{tabular}
\end{ruledtabular}
\end{table*}

\begin{table*}
\caption{
The numbers of structure types whose best structural similarities are isotypic, homeotypic, isopointal, optimal, infeasible, or exceptional, are listed.
Each structure type of metal oxides is analyzed whether it can be the optimal solution of Eq.~\eqref{eq:equation_of_crystal_structures} preserving the experimental structure by assigning the suitable coordination feature to each atom, and the best degree of structural similarity between the experimental and reoptimized structure is determined.
Each structure type is reoptimized after random fluctuation on the positions of atoms, where the maximum displacement size of each atom $\Delta y_i ^{\left( \mathrm{max} \right)}$ is set to be $0$ or $0.01 R_i ^{\left( \mathcal{I} \right)}$.
Note that if an exceptional error such as a negative atomic radius arises during the analysis, the structure is regarded as an exceptional structure.
}
\label{table:matching_rate}
\begin{ruledtabular}
\begin{tabular}{ccc}
Degree of structural similarity & Number of structure types ($\Delta y_i ^{\left( \mathrm{max} \right)} = 0$) & Number of structure types ($\Delta y_i ^{\left( \mathrm{max} \right)} = 0.01 R_i ^{\left( \mathcal{I} \right)}$) \\ \hline
Isotypic & 5077 & 4826 \\
Homeotypic & 778 & 885 \\
Isopointal & 280 & 309 \\
Optimal & 11 & 20 \\
Infeasible & 141 & 302 \\
Exceptional & 188 & 133
\end{tabular}
\end{ruledtabular}
\end{table*}

\subsection{Crystal structure reproduction}

The mathematical optimization problem is applied to several chemical compositions listed in Table \ref{table:target_chemical_compositions}.
The structure types are identified by the structure fingerprints defined in Appendix~\ref{sec:structure_fingerprint}.
Note that largely distorted structures are regarded as different from the symmetric structure.
Many coordination feastures are almost the same as those assigned in the analysis discussed in Sec.~\ref{sec:analysis_on_oxide_crystals_registered_in_ICSD}.
Since cations in a crystal structure have similar cationic repusion radii, we can guess that cations tend to be evenly distributed.
This study also indicate that the number of optimal solutions is much smaller than the number of local minima in the total energy by \textit{ab-initio} simulations~\cite{PhysRevMaterials.8.113801}.

Previous study explicitly constrains the number of common bridging anions as
\begin{equation}
n_{\sigma} ^{\left( \mathrm{common} \right)} \le N_{\sigma} ^{\left( \mathrm{common} \right)},
\label{eq:maximum_number_of_common_bridging_atoms}
\end{equation}
where $N_{\sigma} ^{\left( \mathrm{common} \right)}$ is the maximum number of common bridging anions.
On the other hand, this study only constrains the distances between the cations resulting in the implicit constraints on the connection of coordination polyhedra, which conceptualizes more directly the reason why highly charged cations tend to prefer the vertex sharing to the edge or face sharing, since mathematical crystal chemistry defines interatomic forces as the constraints on interatomic distances.
This alleviation also enables us to automatically set the $N_{\sigma} ^{\left( \mathrm{common} \right)}$ according to the $r^{\left( \mathrm{C} \right)}$ of the chemical elements.
For example, as shown in Fig.~\ref{fig:linked_polyhedra}, two octahedra in \ch{CaCu3Fe4O12} cannot share their edges or faces due to the large $r^{\left(\mathrm{C} \right)}$, but an octahedron and a regular icosahedron can share their faces due to the large ionic bond radius of \ch{Ca}.
Additionally, larger $r^{\left(\mathrm{C} \right)}$ can decreases the number of optimal solutions.
For example, the spinel structure of \ch{Fe2SiO4} shown in Fig.~\ref{fig:iron_oxides}(a) can be unique optimal solution when $r^{\left(\mathrm{C} \right)}$ of silicon atoms is set to be $C^{\left( \mathrm{Pd} \right)}$.
However, \ch{Fe2SiO4} can also constitute the \ch{Mg2SiO4} structure shown in Fig.~\ref{fig:iron_oxides}(b).
This structure, in which a tetrahedron and an octahedron share their edges, can be the optimal solution when $r^{\left(\mathrm{C} \right)}$ of silicon atoms is set to be $C^{\left( \mathrm{E} \right)}$.
Besides, \ch{Fe_{2+$x$}Si_{1-$x$}O4} can constitute the $\beta$-\ch{Mg2SiO4} structure shown in Fig.~\ref{fig:iron_oxides}(c).
This structure can be the optimal solution when $r^{\left(\mathrm{C} \right)}$ of silicon atoms is set to be $C^{\left( \mathrm{O} \right)}$.
In terms of the different compositions, the \ch{CaMgSi2O6}, \ch{Ga2O3}, and \ch{SiFe7O10} structures shown in Figs.~\ref{fig:iron_oxides}(d), (e), and (f), respectively, can be found as the optimal solutions when $r^{\left(\mathrm{C} \right)}$ of silicon atoms is also set to be $C^{\left( \mathrm{O} \right)}$.
This results indicate that too large $r^{\left(\mathrm{C} \right)}$ makes it difficult to reproduce the crystal polymorphism.

The $\alpha$-pyrochlore structure shown in Fig.~\ref{fig:osmium_oxides}(a) can be the unique optimal solution when $R^{\left(\mathrm{I} \right)}$ of mercury atoms, whose coordination number is eight, is set to be $C^{\left( \mathrm{Wd} \right)}$.
On the other hand, the weberite structure shown in Fig.~\ref{fig:osmium_oxides}(b) can be one of the optimal solutions with the $\alpha$-pyrochlore structure when $R^{\left(\mathrm{I} \right)}$ of calcium atoms, whose coordination number is also eight, is set to be $C^{\left( \mathrm{Td} \right)}$.
This result clearly supports the empirical rules of inorganic structural chemsitry that the sizes of atoms are one of the main factors to determine the crystal structures.
On the other hand, $\beta$-pyrochlore structure shown in Fig.~\ref{fig:osmium_oxides}(c) can also be unique optimal solution when $R^{\left(\mathrm{I} \right)}$ of rubidium atoms, whose coordination number is eighteen, is set to be $C^{\left( \mathrm{Qv} \right)}$.
However, although the six ionic bonds of the rubidium is about $0.8$ times smaller than the other 12 ionic bonds, the structure cannot be reproduced if the coordination number of rubidium is set to be six with decreasing $R^{\left(\mathrm{I} \right)}$;
in that case, a rubidium atom makes a largely distorted octahedron resulting in contact with another rubidium atom.
It might be better for alkali metals to alleviate the constraints on the coordination numbers into the constraints that the cation has to be surrounded by anions.

While a calcium ion in the weberite structure constitute the hexagonal bipyramid, the element can constitute a wide variety of coordination polyhedra.
Figure~\ref{fig:calcium_oxides}(a) shows the crystal structure of \ch{Ca2Pt3O8}, and a calcium atom constitutes the triangular prism.
Figure~\ref{fig:calcium_oxides}(b) shows the pyroxene structure of \ch{Ca4Ir4O12}, and a calcium atom constitutes the distorted square antiprism.
Figure~\ref{fig:calcium_oxides}(c) shows the quadruple perovskite structure of \ch{CaCu3Fe4O12}, and a calcium atom constitutes the regular icosahedron.
These results indicate that the mathematical crystal chemistry can reproduce a wide variety of coordination polyhedra resulting in making many kinds of crystal structures.

Previous algorithm was hard to reproduce the square coordinations around \ch{Cu} atoms, but a copper atom in the quadruple perovikite structure makes the square coordination owing to large $r^{\left( \mathrm{C} \right)}$.
This study also finds the crystal structure of \ch{YBa2Cu3O_{7-$x$}} shown in Fig.~\ref{fig:oxide_superconductors}(a).
On the other hand, previous study found the \ch{K2NiF4} structure shown in Fig.~\ref{fig:oxide_superconductors}(b) as the unique optimal solution, but this study additionally finds the distorted \ch{K2NiF4} structure shown in Fig.~\ref{fig:oxide_superconductors}(c) due to small $r^{\left( \mathrm{C} \right)}$.
Finally, the crystal structure of \ch{LaFeAsO} shown in Fig.~\ref{fig:oxide_superconductors}(d) can also be found as unique optimal solution.
Since this structure contains two kinds of anions, we can expect that mathematical crystal chemistry has the capability of crystal structure prediction of not only oxides but also mixed-anion compounds~\cite{10.1039/9781839166372}.

Despite the successful reproduction of crystal structures, computational cost of the algorithm is found to be larger than the previous one.
Since the local structural optimization begins if all the cations just satisfy the constraints on the coordination numbers, a few thousand structural optimization steps waste the time despite the infeasible connections of coordination polyhedra.
Therefore, the inequality constraint of Eq.~\eqref{eq:maximum_number_of_common_bridging_atoms}, which can be determined in advance based on the atomic radii, should be included in Eq.~\eqref{eq:maximization_problem_of_ionic_bonds} to reduce the computational costs as done in the previous study~\cite{PhysRevMaterials.8.113801}.

\subsection{Analysis on oxide crystals registered in ICSD}
\label{sec:analysis_on_oxide_crystals_registered_in_ICSD}

Metal oxides exhibit a wide variety of physical properties such as high-temperature superconductivity and magnetism.
Since ICSD contains a lot of crystal structures of metal oxides, this study analyzes whether they can be the optimal solutions by assigning suitable coordination features to every atom.
First, metal oxides satisfying the five conditions below are collected from the ICSD database.
\begin{itemize}
\item The number of metal atoms is not more than that of oxide ions
\item No anions except for oxide ions
\item No hydrogen atoms
\item No partial occupation of atom sites
\item The number of atoms per unit cell is not more than hundred
\end{itemize}
Second, the different types of them are collected by referencing structure fingerprints.
As a result, $6475$ structure types are discovered.

Since structural reoptimization after the assignment of coordination features transforms the experimental structures, the degree of structural similarity between two structures are defined as listed in Table~\ref{table:definitions_of_matching degree} based on the previous study on inorganic structure types~\cite{https://doi.org/10.1107/S0108767389008834, Allmann:sh0188}.
Note that even if two structures are homeotypic, they do not necessarilly have the same strucure type since in some cases a reoptimized structure is largely deformed.
On the other hand, if the best degree of structural similarity is isopointal, we can expect that the experimental and reoptimized structures have the same structure type, because this case occurs when an experimental structure is too distorted due to such as Jahn-Teller effect;
since the coordination number is underestimated due to the stretched ionic bonds, the experimental and reoptimized structures cannot be isotypic due to the increased coordination numbers of the reoptimized structure.
Finally, if two structures are optimal, in many cases the optimized structure is largely deformed.

The flowchart to search the best degree of structural similarity between experimental and reoptimized structures is shown in Fig.~\ref{fig:matching degree_check_flowchart}.
First, the promising combinations of coordination features are estimated by the algorithm discussed in Appendix~\ref{sec:rule_to_estimate_the_promising_coordination_features_for_analysis_of_oxide_crystals}.
Note that the number of possible combinations of coordination features is too huge due to $25$ kinds of discretized feasible atomic radii listed in Table~\ref{table:atomic_radii}.
Second, a promising combination of coordination features is analyzed whether the experimental structure can be an optimal solution after reoptimization.
If the reoptimized structure is an optimal solution, the experimental structure is additionally reoptimized $20$ times after randomly fluctuating atomic position.
The best degree of structural similarity is determined as the worst one in the $20$ steps.

The result is shown in Table~\ref{table:matching_rate}.
A great majority of the $6475$ kinds of structure type have the optimal solutions preserving the experimental structures whether the random displacements of the atomic positions are applied or not.
When the maximum random displacement size of atomic position is set to be zero, $78\%$, $12\%$, and $4\%$ of them have the isotypic, homeotypic, or isopointal optimal structures, respectively despite the discrete feasible radii listed in Table~\ref{table:atomic_radii}.
The successful results indicate that most of crystal structures of metal oxides can be identified as the optimal solutions of the mathematial optimization problem.
We can expect that not only oxide crystals but also the crystal structures of chalcogenides, halides and mixed-anion compounds~\cite{10.1039/9781839166372} can also be the optimal solutions of the mathematial optimization problem, since they have the structural similarity with oxide crystals.

\section{Conclusion}
\label{sec:conclusion}

Mathematical crystal chemistry formalizes the empirical rules of inorganic structural chemistry as the mathematical optimization problem.
The structural aspects of crystal structures such as linked polyhedra and packing of atomic spheres result from the objective and constraint functions on the optimization problem.
Based on the previous study that defined interatomic forces as the geometrical constraints keeping interatomic distances in stable ranges~\cite{PhysRevMaterials.8.113801}, this study provides a more direct extention of inorganic structural chemistry to refine the theoretical foundation of mathematical crystal chemistry.
First, the minimum and maximum values of the feasible interatomic distance are explicitly defined by the sums of minimum and maximum atomic radii depending on the types of geometrical constraints, respectively, where the geometrical constraints are defined according to the types of interatomic forces based on inorganic structural chemistry.
These radii highlights the aspect that crystal structures are packings of atomic spheres according to multiple types of the atomic radii as shown in Fig.~\ref{fig:concept}.
The definition is an extension of inorganic structural chemistry which estimates a bonding distance by the sum of the effective sizes of atoms depending on the types of bonding forces.
Second, in place of the explicit constraints on the maximum number of common bridging anions, minimum feasible distances between cations implicitly constrain the connection of coordination polyhedra, resulting in the consistency that the constraints on the interatomic distances shape the spatial order of the atoms.
Note that inorganic structural chemistry considers that highly charged cations prefer the vertex sharing to edge or face sharing of coordination polyhedra to reduce the electrostatic repulsion by lengthening the distance between cations.
Third, the creatability functions for the geometrical constraints $\Lambda_{ij \bm{T}} ^{\left( G \right)} \left( \bm{\Phi}, \varepsilon \right)$ similar to activation functions in neural networks are introduced to give a choice of the creatable types of geometrical constraints by inequality constraints.
Optimizing the assignment of the types of geometrical constraints on every pair of atoms depending on the spatial order of the atoms by using $\Lambda_{ij \bm{T}} ^{\left( G \right)} \left( \bm{\Phi}, \varepsilon \right)$ guides a randomly generated structure in the infeasible region toward the optimal solution of the mathematical optimization problem.

The framework makes it possible to unify the paired mathematical optimization problem derived in the previous study~\cite{PhysRevMaterials.8.113801}, resulting in the identification of a wide variety of crystal structures including the spinel, \ch{Mg2SiO4}, $\beta$-\ch{Mg2SiO4}, \ch{Ga2O3}, \ch{CaMgSi2O6}, pyrochlore ($\alpha$ and $\beta$), weberite, \ch{Na2Mg3O8}, pyroxene, quadruple perovskite, \ch{YBa2Cu3O_{7-$x$}}, \ch{K2NiF4}, and \ch{ZrCuSiAs} structures.
Owing to the effectiveness of the strict constraints on the minimum distances between cations, a majority of them can be found as unique optimal solutions, and besides, the square coordination of \ch{Cu} atom is successfully reproduced.
The result shows that the constraints on distances between cations are as dominant as the ionic bonds and the dense packings of anions when modelling ionic crystal structures.
The formalization has some problems such as increased computational cost due to the implicity of the constraints on the connection of coordination polyhedra which can eliminate unpromising networks of ionic bonds.
However, they can explicitly be included in the maximization problem of the number of ionic bonds which is solved every several steps of the global structural optimization in order that random structure search is accelerated as done in the previous study~\cite{PhysRevMaterials.8.113801}.
Furthermore, oxide crystals registered in ICSD are analyzed whether they can be the optimal solutions.
As a result, this study finds that up to $95\%$ of oxide crystal structure types align with the optimal solutions preserving the experimental structures, and at least $83 \%$ of them have the same symmetry and atomic environments despite the discretized feasible atomic radii.

The successful results strongly imply that the mathematical optimization problem deserves to be the theoretical foundation of mathematical crystal chemistry enabling systematic discovery of crystal structures of not only oxides but also such as chalcogenides and mixed-anion compounds with small computations.
Crystal structure prediction can be regarded as the problem how to simultaneously maximize the packing fractions according to the multiple types of the atomic radii, but the creatability functions $\Lambda_{ij \bm{T}} ^{\left( \mathrm{I} \right)} \left( \bm{\Phi}, \varepsilon \right)$ express another aspect that crystal structures are linked polyhedra by creating ionic bonds between suitable pairs of an anion and a cation.
This study can be regarded as the precursor of mathematical crystal chemistry which is the study of how to formalize the empirical rules of inorganic structural chemistry as the objective or constraint functions in mathematical optimization problem.

\appendix

\begin{table*}
\caption{Empirically determined geometrical optimization parameters.}
\label{table:geometrical_optimization_parameters}
\begin{ruledtabular}
\begin{tabular}{cccc}
Parameter & Value for global optimization & Value for local optimization & Value for precise optimization \\ \hline
$S$ & $100$ & $2000$ & $4000$ \\
Initial $\Delta x_s ^{\left( \mathrm{max} \right)}$ & $0.60$ & $0.3$ & $0.1$ \\
Final $\Delta x_s ^{\left( \mathrm{max} \right)}$ & $0.59999$ & $0.05$ & $0.005$ \\
$\varepsilon$ & $1.0$ & $0.2$ & $0.05$
\end{tabular}
\end{ruledtabular}
\end{table*}

\begin{figure}
\centering
\includegraphics[width=1.0\columnwidth]{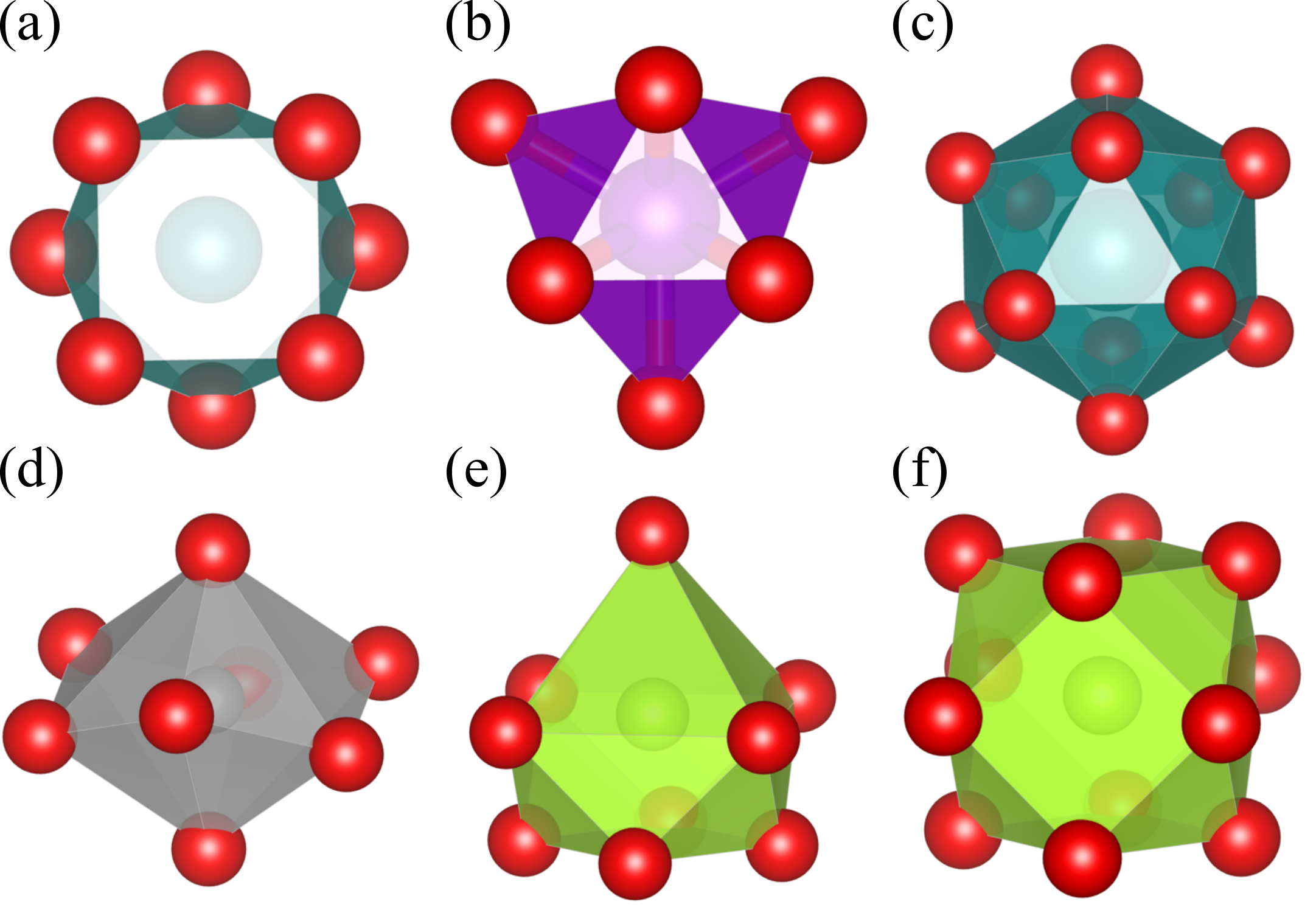}
\caption{
A wide variety of coordination polyhedra can be generated by changing $r^{\left( \mathrm{I} \right)}$, $R^{\left( \mathrm{I} \right)}$, and/or the coordination number.
(a) Square antiprism: The central cation with $r^{\left( \mathrm{I} \right)} = R^{\left( \mathrm{I} \right)} = C^{ \left(\mathrm{E} \right)}$ is surrounded by 8 oxide ions.
(b) Triply-capped trigonal prism: The central cation with $r^{\left( \mathrm{I} \right)} = R^{\left( \mathrm{I} \right)} = C^{ \left(\mathrm{N} \right)}$ is surrounded by 9 oxide ions.
(c) Regular icosahedron: The central cation with $r^{\left( \mathrm{I} \right)} = R^{\left( \mathrm{I} \right)} = C^{ \left(\mathrm{Wd} \right)}$ is surrounded by 12 oxide ions.
(d) Distorted hexagonal bipyramid: The central cation with $r^{\left( \mathrm{I} \right)} = C^{ \left(\mathrm{S} \right)}$ and $R^{\left( \mathrm{I} \right)} = C^{ \left(\mathrm{Wd} \right)}$ is surrounded by 8 oxide ions. This coordination polyhedra can be seen in the $\alpha$-pyrochlore structure shown in Fig.~\ref{fig:osmium_oxides}(a).
(e) Capped square antiprism: The central cation with $r^{\left( \mathrm{I} \right)} = R^{\left( \mathrm{I} \right)} = C^{ \left(\mathrm{O} \right)}$ is surrounded by 9 oxide ions.
(f) Cuboctahedron: The central cation with $r^{\left( \mathrm{I} \right)} = R^{\left( \mathrm{I} \right)} = C^{ \left(\mathrm{O} \right)}$ is surrounded by 12 oxide ions.
}
\label{fig:coordination_polyhedra}
\end{figure}
\begin{figure*}
\centering
\includegraphics[width=2.0\columnwidth]{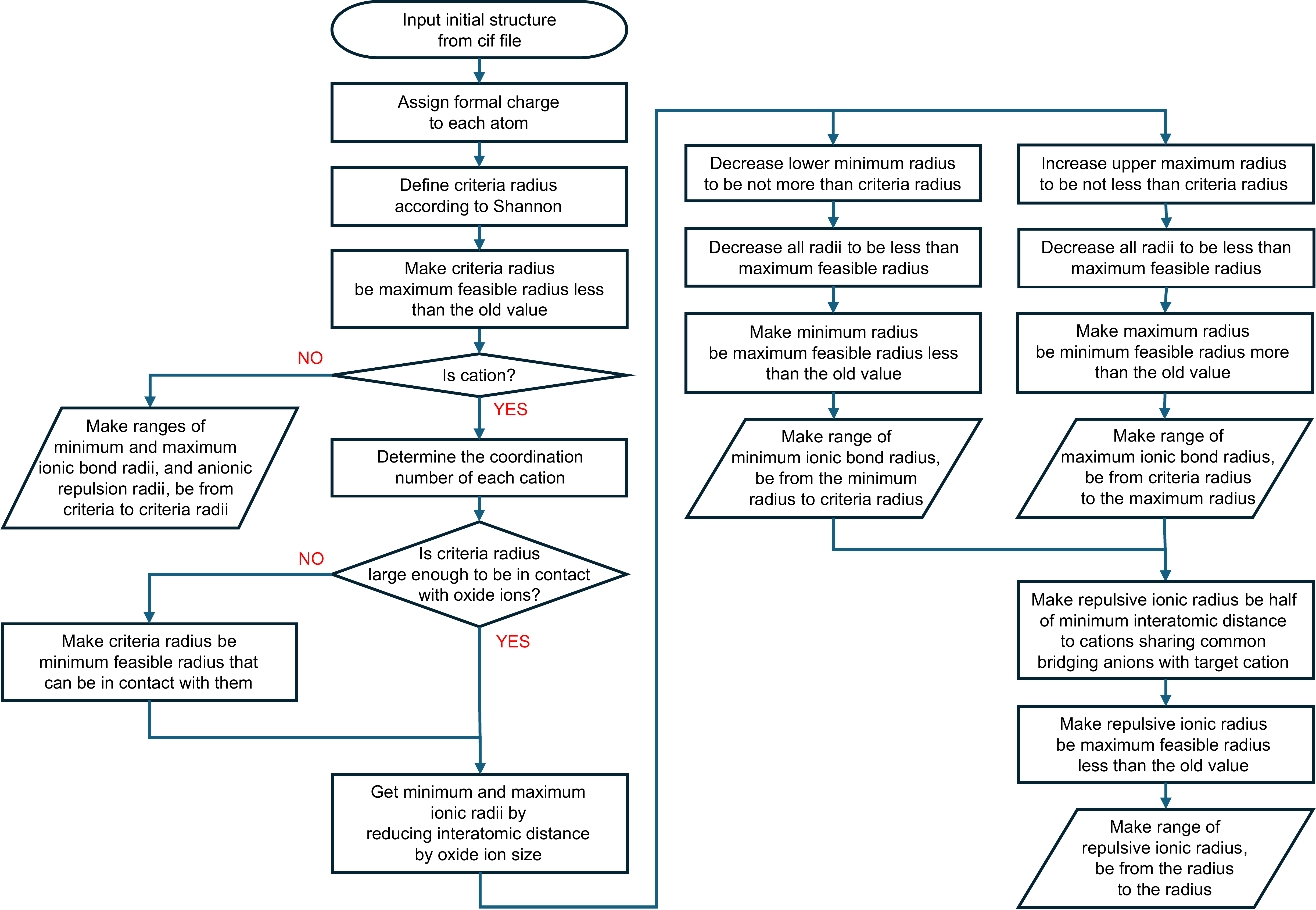}
\caption{
The flow chart of the scheme to estimate the promising combinations of coordination features.
The promising ranges of $r^{\left(\mathrm{I} \right)}$, $R^{\left(\mathrm{I} \right)}$, and $r^{\left(\mathrm{C} \right)}$ of each atom site are assumed based on the ionic radii by Shannon~\cite{SHANNON:a12967}.
}
\label{fig:coordination_feature_search_flowchart}
\end{figure*}

\begin{table}
\caption{
The discretized feasible atomic radii in this study.
The values are the minimum radii so that the central cation can be in contact with all the surrounding $n$ oxygen ions.
The symbols of the feasible radii are determined by the number prefixes derived from from Latin, Greek, or English to avoid same symbols.
}
\label{table:atomic_radii}
\begin{ruledtabular}
\begin{tabular}{ccc}
Symbol & Radius & M.C.N \\ \hline
O & $C^{\left(\mathrm{O} \right)} = 1.400000$ & \\
Z & $C^{\left(\mathrm{Z} \right)} = 0.000000$ & \\
T & $C^{\left(\mathrm{T} \right)} = 0.216581$ & 3 \\
Q & $C^{\left(\mathrm{Q} \right)} = 0.314643$ & 4 \\
P & $C^{\left(\mathrm{P} \right)} = 0.579899$ & 5 \\
H & $C^{\left(\mathrm{H} \right)} = 0.579899$ & 6 \\
S & $C^{\left(\mathrm{S} \right)} = 0.827755$ & 7 \\
E & $C^{\left(\mathrm{E} \right)} = 0.903460$ & 8 \\
N & $C^{\left(\mathrm{N} \right)} = 1.024871$ & 9 \\
D & $C^{\left(\mathrm{D} \right)} = 1.165450$ & 10 \\
Ud & $C^{\left(\mathrm{Ud} \right)} = 1.262958$ & 11 \\
Wd & $C^{\left(\mathrm{Wd} \right)} = 1.262958$ & 12 \\
Td & $C^{\left(\mathrm{Td} \right)} = 1.527604$ & 13 \\
Qd & $C^{\left(\mathrm{Qd} \right)} = 1.598300$ & 14 \\
Pd & $C^{\left(\mathrm{Pd} \right)} = 1.701956$ & 15 \\
Hd & $C^{\left(\mathrm{Hd} \right)} = 1.779744$ & 16 \\
Sd & $C^{\left(\mathrm{Sd} \right)} = 1.846584$ & 17 \\
Ed & $C^{\left(\mathrm{Ed} \right)} = 1.940422$ & 18 \\
Nd & $C^{\left(\mathrm{Nd} \right)} = 2.062955$ & 19 \\
V & $C^{\left(\mathrm{V} \right)} = 2.080892$ & 20 \\
Uv & $C^{\left(\mathrm{Uv} \right)} = 2.21177$ & 21 \\
Wv & $C^{\left(\mathrm{Wv} \right)} = 2.27852$ & 22 \\
Tv & $C^{\left(\mathrm{Tv} \right)} = 2.36085$ & 23 \\
Qv & $C^{\left(\mathrm{Qv} \right)} = 2.36240$ & 24 \\
Pv & $C^{\left(\mathrm{Pv} \right)} = 2.53936$ & 25
\end{tabular}
\end{ruledtabular}
\end{table}

\section{Discretized feasible atomic radii}
\label{sec:discretized_feasible_atomic_radii}

Inorganic structural chemistry considers that all anions surrounding the central cation must be in contact with the central cation for electrostatic stability.
In fact, larger cations in ionic crystals tend to be surrounded by more oxide ions.
Accordingly, the minimum central radius that can be surrounded by $n$ oxide ion is defined as the optimal solution of the optimization problem given by
\begin{equation}
\begin{split}
\text{minimize} \quad & C_0 \\
\text{subject to} \quad & \left| \bm{x}_{j} - \bm{x}_{i} \right| = C_i + C_j
\label{eq:mathematical_programming_for_radii}
\end{split},
\end{equation}
where $C_0$ and $C_k = C^{\left(\mathrm{O} \right)}$ ($k \neq 0$) are the radii of the central cation and oxide ions, respectively.
The optimal solutions of Eq.~\eqref{eq:mathematical_programming_for_radii} are listed in Table~\ref{table:atomic_radii} and the examples of the corresponding coordination polyhedra are shown in Figs.~\ref{fig:coordination_polyhedra}(a), \ref{fig:coordination_polyhedra}(b), and \ref{fig:coordination_polyhedra}(c).
The central cations of the three kinds of polyhedra are the ``hard'' spheres whose minimum and maximum ionic bond radii are the same, but the diversity of coordination polyhedra are enhanced by ``soft'' spheres whose minimum ionic bond radii are less than the maximum ionic bond radii.
For example, the distorted hexagonal bipyramid shown in Fig.~\ref{fig:coordination_polyhedra}(d) can be generated by an atom whose minimum and maximum ionic bond radii are $C^{\left(\mathrm{S} \right)}$ and $C^{\left(\mathrm{Wd} \right)}$, respectively.
Besides, even if the coordination numbers are the same, two hard spheres having the different radii can generate different coordination polyhedra.
For example, the hard cation whose ionic bond radii is $C^{\left(\mathrm{Wd} \right)}$ generates the regular icosahedron shown in Fig.~\ref{fig:coordination_polyhedra}(b), while the cation whose ionic bond radii is $C^{\left(\mathrm{O} \right)}$ generates the cuboctahedron shown in Fig.~\ref{fig:coordination_polyhedra}(f).
On the other hand, even if the sizes of the two ``hard'' spheres are the same, they can generate different coordination polyhedra if the coordination numbers are different.
For example, both the capped square antiprism shown in Fig.~\ref{fig:coordination_polyhedra}(e) and cuboctahedron shown in Fig.~\ref{fig:coordination_polyhedra}(f) can be generated by ``hard'' spheres whose ionic bond radii are $C^{\left(\mathrm{O} \right)}$.
In summary, only the three simple parameters of the minimum ionic bond radius, maximum ionic bond radius, and the coordination number enable us to generate a wide variety of coordination polyhedra.
In this study, all the atomic radii utilized in this study are discretized so that they have to be equal to one of the feasible radii.

\section{Initial structure generation}
\label{sec:initial_structure_generation}

An initial structure is randomly generated as follows:
Let $\left( a,b,c, \alpha, \beta, \gamma \right)$ be the lattice parameters, and they are initialized randomly under the condition of
\begin{align}
1 \le b,c \le 3, &&
\frac{1}{3} \pi \le \alpha, \beta, \gamma \le \frac{2}{3} \pi,
\end{align}
where $a$ is set to be $1$.
Next, lattice vectors $\left( \bm{t}_1, \bm{t}_2, \bm{t}_3 \right)$ is scaled to satisfy the condition of
\begin{equation}
\frac{1}{\left| \bm{t}_1 \cdot \left( \bm{t}_2 \times \bm{t}_3 \right) \right|} \left[\sum_{i=1} ^M \frac{4}{3} \pi \left( R_i ^{\left( \mathrm{I} \right)} \right)^3 \right] = 0.7.
\end{equation}
Finally, all the atoms are randomly distributed in the unit cell.

\section{Structural optimization algorithm}
\label{sec:structural_optimization_algorithm}

The structural optimization algorithm devised in the previous study~\cite{PhysRevMaterials.8.113801} is summarized to present the paper in a self-contained manner.
The inequality constraints are approximated by the hard-spherical potential as
\begin{align}
d_{\sigma} \le x_{\sigma} \; &\Rightarrow \; U_{\mathrm{min}} \left(x_{\sigma} \right) \equiv \max \left[ 0, k_{\downarrow} \left( d_{ij} - x_{\sigma} \right) \right], \\
x_{\sigma} \le D_{\sigma} \; &\Rightarrow \; U_{\mathrm{max}} \left(x_{\sigma} \right) \equiv \max \left[ 0, k_{\uparrow} \left( x_{\sigma} - D_{ij} \right) \right],
\end{align}
where $k_{\downarrow}$ and $k_{\uparrow}$ are a common constant for repulsive and attractive forces, respectively.
Accordingly, the structural optimization problem is given by
\begin{equation}
\text{minimize} \quad H \left( \bm{X} \right),
\end{equation}
where the objective function $H \left( \bm{X} \right)$ is defined as
\begin{equation}
H \left( \bm{X} \right) \equiv \sum_{\sigma} \left[ U_{\mathrm{min}} \left(x_{\sigma} \right) + U_{\mathrm{max}} \left(x_{\sigma} \right) \right] + P \Omega,
\end{equation}
with $P$ being the pressure.
The minimization problem is solved by iterative-balance methods~\cite{PhysRevE.103.023307, PhysRevE.104.024101, koshoji2021diverse, PhysRevMaterials.8.113801} as follows:
Let $\Delta \bm{x}_i$ and $\Delta \bm{t}_i$ be the displacements of $\bm{x}_i$ and $\bm{t}_i$ calculated by
\begin{align}
\Delta \bm{x}_i = - \xi_i \frac{\partial H \left( \bm{X} \right)}{\partial \bm{x}_i}, && \Delta \bm{t}_i = - \zeta_i \frac{\partial H \left( \bm{X} \right)}{\partial \bm{t}_i},
\end{align}
where the constants $\xi_i$ and $\zeta_i$ are scaled to satisfy the condition given by
\begin{align}
\left| \Delta \bm{x}_i \right| \le \Delta x_s ^{\left( \mathrm{max} \right)}, && \left| \Delta \bm{t}_i \right| \le \gamma \Delta x_s ^{\left( \mathrm{max} \right)},
\end{align}
with $\Delta x_s ^{\left( \mathrm{max} \right)}$ being the maximum displacement of atoms in $s$-th optimization step and $\gamma$ being a constant.
$\Delta x_s ^{\left( \mathrm{max} \right)}$ is calculated as
\begin{equation}
\Delta x_s ^{\left( \mathrm{max} \right)} = \Delta x_0 ^{\left( \mathrm{max} \right)} \left( \frac{\Delta x_{S} ^{\left( \mathrm{max} \right)}}{\Delta x_0 ^{\left( \mathrm{max} \right)}} \right) ^{\frac{s}{S}},
\end{equation}
where $S$ is the maximum number of optimization steps, and $\gamma$ is defined as
\begin{equation}
\gamma = 0.02 L \cdot \Delta x_s ^{\left( \mathrm{max} \right)},
\end{equation}
with $L$ being the number of atoms per unit cell.
In this study, the force constants are set to be
\begin{align}
P &= 1.0, && k_{\uparrow} = 30.0, && k_{\downarrow} = -100.0.
\end{align}
In the global optimization, $\bm{n}$ is optimized every $20$ structural optimization steps, and the unit cell is refined per 1000 steps by using the software SPGLIB~\cite{Togo31122024}.
In the local optimization, if a structure cannot be an optimal solution after the first $2000$ steps, the structure is optimized again after initializing $\Delta x_s ^{\left( \mathrm{max} \right)}$, because in some cases, not all the interatomic distances converge into the feasible regions.
The other parameters used in the structural optimization are listed in Table~\ref{table:geometrical_optimization_parameters}.
Note that the error rate $\varepsilon$ is set to be zero throughout the structural optimization.

The cutoff distances of the ionic bonds are set to be $2 D_{\sigma} ^{\left( \mathrm{I} \right)}$ to reduce the computational cost.
Similarly, the cutoff radii of the non-ionic bonds, anionic constraints, cationic constraints, and metallic constraints are set to be $2 d_{\sigma} ^{\left( G \right)}$.

\section{Structure fingerprint}
\label{sec:structure_fingerprint}

The previous study uses several criterias such as the space group types, Wyckoff sequence, $c/a$ ratio, and $\beta$ ranges to identify the inorganic structure types~\cite{https://doi.org/10.1107/S0108767389008834, Allmann:sh0188}.
In many cases, the space group types and Wyckoff sequence are enough to distinguish two crystal structures.
Based on the insight into the inorganic crystal structures, the structure fingerprint (SF) is defined by not only the symmetry of the crystal but also the geometrically constrained state.
SF is specified by the space group number and the list of the atomic environments (AEs), where AE is defined by the subset the atom belongs, the site symmetry symbol, the number of ionic bonds, and the connection of polyhedra which can be recognized from the geometrical constraints.
Additionally, the simplified structure fingerprint (SSF) is defined as the list of simplified atomic environment which is defined as the AE without the site symmetry symbol.
In this study, the space groups are determined by using SPGLIB~\cite{Togo31122024}.

\section{Rule to estimate the promising coordination features for analysis of oxide crystals}
\label{sec:rule_to_estimate_the_promising_coordination_features_for_analysis_of_oxide_crystals}

In the analysis on the oxide crystals registered in ICSD, the promising ranges of $R^{\left(\mathrm{I} \right)}$, $r^{\left(\mathrm{I} \right)}$, and $r^{\left(\mathrm{C} \right)}$ of each atom site are assumed following the flowchart shown in Fig.~\ref{fig:coordination_feature_search_flowchart}.
The formal charge of each atom is determined so that the sum of ionization potentials are minimized:
\begin{equation}
\begin{split}
\text{minimize} \quad & \sum_{i=1}^M \sum_{n=0}^{q_i} I_i ^{\left( n \right)} \chi_i ^{\left(\mathcal{M} \right)} \\
\text{subject to} \quad & q_i ^{\left( \mathrm{min} \right)} \le q_i \le q_i ^{\left( \mathrm{max} \right)} \\
& \left| q_i - q_j \right| \, \chi_i ^{\left(\mathcal{A} \right)} \chi_j ^{\left(\mathcal{A} \right)} \le 1
\end{split},
\label{eq:minimization_problem_of_ionic_potentials}
\end{equation}
where $I_i ^{\left( n \right)}$ is the experimentally determined $n$-th ionization potential~\cite{IntroductionToStructuralChemistry}, $q_i$ is the formal charge of atom $i$, and $q_i ^{\left( \mathrm{min} \right)}$ and $q_i ^{\left( \mathrm{max} \right)}$ are the minimum and maximum formal charges of atom $i$, respectively.
Note that the second constraint makes the anions be evenly reduced.

If the number of the combinations of the coordination features derived from the algorithm shown in Fig.~\ref{fig:coordination_feature_search_flowchart} is more than $100000$, the ranges of $r^{\left(\mathrm{I} \right)}$ is limited to the criteria radius defined in Fig.~\ref{fig:coordination_feature_search_flowchart}.
If it does not work, the ranges of $R^{\left(\mathrm{I} \right)}$ is also limited to the criteria radius defined in Fig.~\ref{fig:coordination_feature_search_flowchart}.
On the other hand, if the number of the combinations is not more than $100000$, the ranges of $R^{\left(\mathrm{I} \right)}$ and $r^{\left(\mathrm{C} \right)}$ are extended as follows:
First, the range of $R^{\left(\mathrm{I} \right)}$ is extended so that the maximum coordiation numbers can be two larger unless $C^{\left( \mathrm{Pv} \right)} < R^{\left(\mathrm{I} \right)}$.
Second, the range of $r^{\left( \mathrm{C} \right)}$ is extended so that it contains one additional feasible radius which is lowest in the feasible radii more than the criteria radius.
However, these extentions of the ranges are stopped if the number of the combinations of the coordination features is more than $100000$.
Finally, $r^{\left( \mathrm{M} \right)}$ of each cation is set to be $C^{\left( \mathrm{O} \right)}$.

\begin{acknowledgements}

This research was conducted using FUJITSU Server PRIMERGY GX2570 (Wisteria/BDEC-01) at the Information Technology Center, The University of Tokyo, and the facilities (supercomputer Ohtaka) of the Supercomputer Center, the Institute for Solid State Physics, the University of Tokyo.
The author would like to thank Prof.~Taisuke Ozaki, Takeshi Yajima, Daigorou Hirai, Jun-ichi Yamaura, and Yoshihiko Okamoto for kindly discussing on mathematical crystal chemistry.
Finally, the author also would like to thank Profs.~Satoru Iwata and Yoshihiro Kanno for discussing on mathematical aspects.

\end{acknowledgements}

\bibliography{draft9}

\end{document}